\documentclass[english,submission]{programming}

\usepackage[backend=biber]{biblatex}
\usepackage{amsthm}
\theoremstyle{definition}
\newtheorem*{remark}{Remark}
\newtheorem*{note}{Note}
\newtheorem{casestudy}{Case study}

\usepackage{mathtools}
\usepackage{csquotes}
\usepackage{stmaryrd}
\usepackage{quiver}
\usepackage{hhline}

\usepackage{circledsteps}
\newcommand{\circledtext}[1]{{\sffamily\Circled{#1}}}

\newcommand{\citet}[1]{\citeauthor*{#1}~\cite{#1}}

\newcommand{\mathbox}[1]{\colorbox{black!10}{$#1$\phantom{i\hspace{-3.5pt}}}}
\usepackage{tcolorbox}
\tcbset{colback=black!10, colframe=white, right=-20pt}
\newcommand{\figbox}[1]{\tcbox{
  \begin{minipage}{\textwidth}
  \vspace{-14pt}
  #1
  \end{minipage}}}

\DeclareMathOperator{\exec}{\ \fatsemi\ \ }
\newcommand{\is}{{:}\ }
\newcommand{\comma}{,\ }
\newcommand{\isa}{\dblcolon}
\newcommand{\quotedstring}[1]{\textsf{\textquotedbl{#1}\textquotedbl}}
\newcommand{\emptystring}{\textsf{\textquotedbl\ \!\textquotedbl}}

\newcommand{\myparagraph}[1]{\vspace{-8pt}\paragraph{#1}}

\paperdetails{
  perspective=art,
  area={Database programming, Visual and live programming, Programming environments, Schema evolution, Version control},
}

\begin{document}

\title{Baseline: Operation-Based Evolution and Versioning of Data}

\author[a]{Jonathan Edwards}[0000-0003-1958-7967]
\authorinfo{is an independent researcher interested in simplifying and democratising programming by collapsing the tech stack. He is known
for his \href{https://subtext-lang.org}{Subtext} programming language experiments and his blog at \href{https://alarmingdevelopment.org}{\texttt{alarmingdevelopment.org}}.
He has been a researcher at MIT CSAIL and CDG/HARC. He tweets \href{https://x.com/jonathoda}{\texttt{@jonathoda}} and can be reached at
\email{jonathanmedwards@gmail.com}.}
\affiliation[a]{Independent, Boston, MA, USA}

\author[b]{Tomas Petricek}[0000-0002-7242-2208]
\authorinfo{is an assistant professor at Charles University. He
is interested in finding easier and more accessible ways of thinking
about programming. To do so, he combines technical work on
programming systems and tools with research into history and
philosophy of science. His work can be found at \href{https://tomasp.net}{\texttt{tomasp.net}} and
he can be reached at \email{tomas@tomasp.net}.}
\affiliation[b]{Charles University, Prague, Czechia}

\keywords{Schema evolution, version control, operational transformation} 

\ccsdesc[500]{Software and its engineering~Software configuration management and version control systems}
\ccsdesc[500]{Information systems~Query languages}
\ccsdesc[500]{Information systems~Database utilities and tools}

\maketitle

\begin{abstract}

  Baseline is a platform for richly structured data supporting change in multiple dimensions: mutation over time, collaboration across space, and evolution through design changes. It is built upon \textit{Operational Differencing}, a new technique for managing data in terms of high-level operations that include refactorings and schema changes. We use operational differencing to construct an operation-based form of version control on data structures used in programming languages and relational databases.

  This approach to data version control offers high-fidelity diffing and merging despite intervening structural transformations like schema changes. It offers users a simplified conceptual model of version control for ad hoc usage: There is no repo; Branching is just copying. The information maintained in a repo can be synthesized more precisely from the append-only histories of branches. Branches can be flexibly shared as is commonly done with document files, except with the added benefit of diffing and merging.

  We conjecture that queries can be \textit{operationalized} into a sequence of schema and data operations. We develop that idea on a query language fragment containing selects and joins.
  Operationalized queries are represented as a \textit{future timeline} that is speculatively executed as a branch off of the present state, returning a value from its \textit{hypothetical future}. Operationalized queries get rewritten to accommodate schema change ``for free'' by the machinery of operational differencing.


  We evaluate Baseline with case studies of schema evolution problems identified iin a literature review, including the set of challenge problems we identified in a prior paper.


\noindent\rule{\textwidth}{1pt}

Context: Many technologies have been developed to manage data, yet we still lack convenient general purpose tools for schema evolution and version control of data.

Inquiry: Can the problems of schema evolution and data version control be solved with an operation-based technique? We build upon research on Schema Modification Operations and Operational Transformation.

Approach: Most tools for data management are state-based, for pragmatic reasons of interoperability. We are exploring what is possible with an operation-based approach.

Knowledge: Operational Differencing is able to do fine-grained version control of data even through intervening schema changes. It can be used to synthesize more precisely the information traditionally maintained in a repo. Version control without a repo is simpler and more flexible for ad hoc usage.

Grounding: We evaluate Baseline with case studies of schema evolution problems identified iin a literature review, including the set of challenge problems we identified in a prior paper.

Importance: Our technique improves upon the known methods for schema evolution and version control of data. Our conjecture that queries can be operationalized could open up a new space of language designs.

\end{abstract}

\section{Introduction}

Data changes in multiple dimensions. Data changes over time, indeed often that is the whole point. Data changes over space between replicas and variant copies. Data also changes in meaning: as needs and understanding evolve so must the shape and interpretation of the data. In practice we handle these different dimensions of change with different technologies embodying different abstractions. Programming languages, file systems, and databases all have distinct ways of managing data change over time. There are many different techniques for distributing, replicating, and collaborating on data, although surprisingly few for version control of data. There is a plethora of tools to assist in evolving the schema of databases, but in practice it often requires ad hoc programming.

Baseline is a research project to develop a general purpose mechanism for data change in all these dimensions, and to provide it conveniently to both programs and users.
To ground this research we first surveyed one of the most problematic aspects of data change: schema evolution. In \textit{Schema Evolution in Interactive Programming Systems}~\cite{challenge-problems} we and others formulated a suite of challenge problems across a range of contexts, including traditional database programming as well as live front-end programming, model-driven development, and collaboration in computational documents.

Systems that manage change split into \textit{state-based} and \textit{operation-based} approaches~\cite{diff3, Shapiro11}. State-based approaches observe only the before and after of changes, while operation-based approaches record the execution of a set of possible operations.
The benefit of state-based approaches is that since they observe from the outside they more easily interoperate with existing tools.
The benefit of operation-based approaches is that they can capture user intentions more accurately than can be inferred from the states. For example comparing states of a table schema, a move followed by a rename is indistinguishable from a delete followed by an insert, but that makes a big difference to the data.

In this paper we introduce a new technique for operation-based change management: \textit{Operational Differencing}, which we define on a \textit{rich data} model curated from the features of programming languages and databases.


Baseline manifests all operations in a direction manipulation UI that gives users the same power as programs.
It is difficult to convey the experience of an interactive interface in a written narrative, so we offer the reader a playable demo at \url{https://thebaseline.dev/Prog26submission/}.
Space limitations force us to move discussion of the UI and the demo into Appendices \ref{UI} and \ref{demo}.

The contributions of this paper are:

\begin{enumerate}

\item We introduce \textit{Operational Differencing} by example in the familiar setting of typed nested lists and records supplied with a rich set of operations
including refactorings. We define the primitive transformations \textit{projection} and \textit{retraction} with rules expressed as commutative diagrams.

\item Using these primitives we construct the key capabilities of version control: \textit{diffing} and \textit{transfer} (a merging primitive somewhat like git's cherry-pick). Being operation-based provides high-fidelity diffing and merging across refactorings and structural transformations. For example renaming doesn't conflict with other changes, and movement of a structure doesn't conflict with changes within it.

\item We propose a simplified model of version control without a repo: documents have an append-only history of operations; branching is just making a copy of the artifact with its history, which can then diverge. The graph of branches and merges does not need to recorded in a repo because it can be synthesized from the histories on need, and can be done so optimally in a sense we define. Liberating versions and branches from the repo frees them to be shared independently, as in the ubiquitous folk practice of copying spreadsheets~\cite{Burnett14, Basman19}, but with the added benefit of being able to diff and merge.

\item We enrich the data model to incorporate relational databases, including operations for schema change and normalization, providing version control on databases that works across schema changes and automates schema migration.

\item We conjecture that queries can be \textit{operationalized} into a sequence of schema and data operations, in a form of Programming by Demonstration~\cite{cypher93-pbd}. We develop this idea on a query language fragment with selects and joins.
Query rewriting~\cite{curino08, herrmann17} during schema change falls out ``for free''.



\end{enumerate}

We evaluate the capabilities and usefulness of Baseline with case studies of schema evolution scenarios taken from the research literature. These include the challenge problems we identified earlier~\cite{challenge-problems}.


\section{Rich Data Operations}\label{rich-data}

We start with familiar data structures: nested lists and records. The atomic values are strings, supplied in the set \mathbox{S}, and numbers, supplied in \mathbox{N}, including \mathbox{\textsf{NaN}}. What is unusual about this data model is that we assign permanent unique identifiers (IDs) to every record field and list element. These IDs are supplied from the disjoint sets \mathbox{F} for record fields and \mathbox{E} for list elements. Because record fields have unique IDs their names are only for human readability and we elide them in most examples. Another unusual feature is that deletion of a record field or list element leaves behind a \textit{tombstone} (see Appendix~\ref{deletion}).


The data model is typed similarly to statically typed Functional Programming (FP) language such as ML, as follows. List elements all have the same type as defined in the type of the list (tombstones are a bottom type). Record values have the same sequence of fields with the same ID and type of value as in the type of the record. The elements of a list and the fields of a record must have different IDs.
The empty record \mathbox{\text{\{\}}}serves as a unit type. Every type has an \textit{initial value}.
We omit sum types and type aliases because they are not needed in this paper.
Figure~\ref{rich-data} defines the syntax and operations of the model.

\begin{figure}[h]
  \vspace{-8pt}
\figbox{
\[ \begin{array}{r@{\ }l|r@{\ }l|r@{\ }l|l}
  \multicolumn{2}{l|}{\textrm{type}} & \multicolumn{2}{l|}{\textrm{value}} & \multicolumn{2}{l|}{\textrm{initial value}}&\\
  \hline
  T \Coloneqq & & v \Coloneqq & & T^\varnothing = & &\\
  &  \textsf{String} & & S & & \emptystring & \textrm{string}\\
  & \textsf{Number} & &  N & & \textsf{NaN} & \textrm{number}\\
  & \textsf{List } T & & [ E \is v \  \dots ] & & [] & \textrm{list}\\
  & \{ F \  S \is T \  \dots \} & & \{ F \is v \  \dots \} & & \{ F \is T^\varnothing \  \dots \}& \textrm{record}\\

  & \bot && \bigtimes & & \bigtimes & \textrm{tombstone}\\
\end{array}\]

\vspace{-12pt}
\[
\hspace{-22pt}
\begin{array}{l|l}
  \textrm{Value operations} & \\
  \hline
  p \textsf{ noop} & \textrm{Do nothing}\\
  p \textsf{ write } v & \textrm{Write atomic value $v$ at $p$}\\

  p \textsf{ insert } E \textsf{ before } E' & \textrm{Insert element ID $E$ into list at $p$ in front of element $E'$}\\
  p \textsf{ append } E & \textrm{Append element ID $E$ to end of list at $p$}\\

  p \textsf{ delete } E & \textrm{Delete element ID $E$ in list at $p$}\\

  p \textsf{ move } p' & \textrm{List element $p$ is copied to $p'$ in same list then $p$ is deleted}\\
  \hline

  \textrm{Type operations} & \rule[0pt]{0pt}{16pt}\\
  \hline
  p \textsf{ Define } T & \textrm{Define $p$ to have type $T$, initializing values}\\
  p \textsf{ Convert } T & \textrm{Convert atomically-typed $p$ to atomic type $T$}\\

  p \textsf{ Rename } S & \textrm{Rename record field $p$ to string $S$}\\

  p \textsf{ Insert } F \textsf{ before } F' & \textrm{Insert field ID $F$ into record at $p$ in front of field $F'$}\\
  p \textsf{ Append } F & \textrm{Append field ID $F$ to end of record at $p$}\\

  p \textsf{ Delete } F & \textrm{Delete field ID $F$ in record at $p$}\\

  p \textsf{ Move } p' & \textrm{Record field at $p$ is copied to $p'$ in the same record}\\
  & \textrm{or containing\slash contained record, and then $p$ is deleted}\\

  p \textsf{ ListOf } & \textrm{Convert value at $p$ into a list of one element with ID $\mathbb{1}$}\\
  p \textsf{ IntoFirst } & \textrm{Convert list at $p$ into its first element, else the initial value} \\

  p \textsf{ RecordOf } F & \textrm{Convert value at $p$ into a record of one field with ID $F$} \\
  p \textsf{ IntoField } F & \textrm{Convert record at $p$ into value of its field $F$} \\

\end{array}\]
}
\vspace{-4pt}
\caption{Rich data syntax and operations}
\label{fig:rich-data}
\end{figure}

A \textit{path} \mathbox{p} is a possibly empty sequence of IDs denoting a path drilling into nested records and lists. Paths can access both values and types. We separate the IDs in a path with dots, and a single dot is the empty (top) path. A special element ID~\mathbox{*} is used to access the element type of a list type.

A \textit{state} pairs a value with its type. We abuse the notation \mathbox{v \isa T} to both denote a state and assert that the value matches the type.
An \textit{operation} is a partial function from an \textit{input state} to an \textit{output state}. Operations are not defined on all input states, imposing various preconditions.
All operations take a \textit{target} path as a parameter indicating that the operation is to be performed at that path within the state.
Operations use infix syntax, placing the target path first, then the operation name, followed by any other parameters.


We execute operations with the \mathbox{\fatsemi} infix operator taking an input state on the left, an operation on the right, and yielding an output state which may be chained into subsequent executions. Here is an example that starts with an empty list of numbers and adds an element with ID $e$ and value $1$:
\begin{align*}
  S_1 &= [] \isa \textsf{List Number}\\
  S_2 &= S_1 \exec \ldotp \textsf{ append } e \exec \ldotp e \textsf{ write } 1\\
  &= [e \is 1] \isa \textsf{List Number}\\
\end{align*}
\vspace{-30pt}\\
The \textsf{append} targets the empty path \mathbox{\ldotp} referring to the whole list. Then the \textsf{write} targets the new element with the path \mathbox{\ldotp e}. In all our examples we take IDs like \mathbox{e} as given but in practice the Baseline API generates them uniquely.

A \textit{table} is a list of records, the fields of which are the \textit{columns} while the elements of the list are the \textit{rows}. Here is an example of changing the type of a table:
\begin{align*}
S_1 &= [e \is \{x \is 1 \comma  y \is 2\}] \isa \textsf{List} \{x \is \textsf{Number}\comma  y \is \textsf{Number}\}\\
S_2 &= S_1 \exec \ldotp * \textsf{ Append } z \exec \ldotp * \ldotp z \textsf{ Define Number}\\
 &= [e \is \{x \is 1\comma  y \is 2,\ z\is \textsf{NaN}\}] \isa \textsf{List} \{x \is \textsf{Number}\comma  y \is \textsf{Number}\comma z\is \textsf{Number}\}\\
\end{align*}
\vspace{-30pt}\\
The \textsf{Append} operation targets the path \mathbox{\ldotp *} which is the record type of the list's elements and adds a new field \mathbox{z} to it, in effect adding a column to the table. The \textsf{Define} operation targets the path \mathbox{\ldotp  * \ldotp z} which is the type of the new field (initially the unit type \mathbox{\{\}}) and changes it to \mathbox{\textsf{Number}}. The new field is also inserted into all elements of the list with the initial value \mathbox{\textsf{NaN}}.

The \textsf{append} and \textsf{write} operations are \textit{value operations}, meaning they do not modify the type of the state. On the other hand \textsf{Append} and \textsf{Define} are \textit{type operations} that may modify both the type and value. Type operations do \textit{schema migration}: the value is adapted to match the type while minimizing loss of information, iterating over list elements as needed (think of the \mathbox{*} in the type path as a wildcard). By convention we capitalize the name of type operations. Baseline has a \textit{user mode} in which only value operations are permitted.

\begin{remark}
In this presentation we adopt the standard approach of defining values and types as distinct mathematical objects. Actually Baseline only has values, with initial values serving as prototypes. Every list contains an initial \textit{header} element with the ID~\mathbox{*} containing an initial value which serves as the prototype of the list elements. Note how this convention corresponds to the visual rendering of a table where a header row describes the type of each column. Types are a powerful abstraction for the theory and implementation of programming systems but we conjecture that they can be replaced with prototypes, without loss of power, to simplify the programming experience. We have also explored an untyped version of our approach~\cite{denicek}.
\end{remark}

Many of the type operations exist in order to do schema migration on values: they would not be needed just to define a type prior to populating it with data. We think of these operations as \textit{type refactorings}, capturing high-level design changes. They are manifested in the UI as direct manipulations so that schema evolution can be done interactively. Interactive schema evolution is also needed in \textit{live programming}~\cite{challenge-problems}.

\begin{figure}[]
  \vspace{-20pt}
\begin{align*}
  S_1 &= [
    e \is \{ \mathit{what}\is \quotedstring{clean}\comma  \mathit{who}\is \quotedstring{Jack} \}
    ] \isa
    \textsf{List}\{ \mathit{what}\is \textsf{String}\comma \mathit{who}\is \textsf{String} \}\\
S_2 &= S_1 \exec \ldotp *\ldotp \mathit{who} \textsf{ ListOf } \\
 &= [
    e \is \{ \mathit{what}\is \quotedstring{clean}\comma  \mathit{who}\is [\mathbb{1}\is \quotedstring{Jack}] \}
    ] \isa
    \textsf{List}\{ \mathit{what}\is \textsf{String}\comma  \mathit{who}\is \textsf{List String} \}\\
\end{align*}
\vspace{-40pt}
\caption{TODO refactoring}
\label{fig:TODO-refactor}
\end{figure}

For example a common design evolution is when a single value needs to become a list of multiple values. We capture this in the \textsf{ListOf} operation, shown on a table of tasks for a TODO application in Figure~\ref{fig:TODO-refactor}. We want to change tasks from having a single assigned person to a list of people.
The \textsf{ListOf} operation takes any type and wraps it inside a list, and wraps all corresponding values into single-element lists using the element ID $\mathbb{1}$. We will return to this example in the next section.


\section{Operational Differencing}

Change management systems deal with what happens when two copies (or replicas or branches) of something diverge.
These systems split into \textit{state-based} and \textit{operation-based} approaches~\cite{diff3, Shapiro11}.
In this section we introduce a new form of operation-based change management: \textit{Operational Differencing}.

The benefit of operation-based techniques is that they can capture more information than can be inferred from the states.
For example a \textsf{Rename} operation makes many changes throughout the state, and standard version control systems will see these changes as unrelated and conflicting with any other changes on the same text lines. In contrast operational differencing sees a renaming as a single change that conflicts only with renaming to a different name.
On the other hand, the benefit of state-based techniques is that they can observe other systems from the outside, which may explain why all existing version control systems are state-based.

\begin{remark}
We discuss source code version control systems like git~\cite{ProGit} as a point of reference, but we do not propose to compete with them. Git is highly adapted to the needs of intensive development in the Unix environment and the cultural norms of Unix developers. In that context version control must be state-based in order to interoperate with existing tools and practices. Instead of source code we focus on rich data, where we believe operation-based version control offers important benefits.
\end{remark}

We define a \textit{timeline} to be an initial state and a sequence of operations executed in order that are each valid on their input state, producing an end state.
A \textit{history} is a timeline starting from the initial empty state of a document, which by convention is the unit value.
Baseline records histories by observing the operations executed in the API and in a UI that manifests operations as direct manipulations. Operational differencing is built upon two primitive functions on operations: \textsf{Project} and \textsf{Retract}.

\subsection{Projection}

\begin{figure}[h]
  \vspace{-8pt}
\begin{minipage}[b][][b]{1.75in}
\center
\begin{tikzcd}[column sep=large]
	0 & {[\mathbb{1}\is 0]} \\
	2 & {[\mathbb{1}\is 2]}
	\arrow["{{\ldotp \textsf{ ListOf}}}", from=1-1, to=1-2]
	\arrow["{{\ldotp \textsf{ write } 2}}"'{pos=0.4}, from=1-1, to=2-1]
	\arrow["{{\textsf{P}\Rightarrow}}"{description, pos=0.2}, draw=none, from=1-1, to=2-2]
	\arrow["{{\ldotp\mathbb{1} \textsf{ write } 2}}"{pos=0.4}, dashed, from=1-2, to=2-2]
	\arrow["{{\ldotp \textsf{ ListOf}}}"', dashed, from=2-1, to=2-2]
\end{tikzcd}
\\[4pt]
(a)
\end{minipage}
\quad
\begin{minipage}[b][][b]{1.75in}
\center
\begin{tikzcd}[column sep=large]
	0 & 1 \\
	2 & 2
	\arrow["{{\ldotp \textsf{ write } 1}}", from=1-1, to=1-2]
	\arrow["{{\ldotp \textsf{ write } 2}}"'{pos=0.4}, from=1-1, to=2-1]
	\arrow["{{\textsf{P}\Rightarrow}}"{description, pos=0.2}, draw=none, from=1-1, to=2-2]
	\arrow["{{\ldotp \textsf{ write } 2}}"{pos=0.4}, dashed, from=1-2, to=2-2]
	\arrow["{{\ldotp \textsf{ noop}}}"', dashed, from=2-1, to=2-2]
\end{tikzcd}
\\[4pt]
(b)
\end{minipage}
\quad
\begin{minipage}[b][][b]{1.25in}
\center
\begin{tikzcd}[column sep=large]
	\bullet & \bullet \\
	\bullet & \bullet
	\arrow["{\mathit{base}}", from=1-1, to=1-2]
	\arrow["{\mathit{pre}}"'{pos=0.4}, from=1-1, to=2-1]
	\arrow["{\textsf{P}\Rightarrow}"{description, pos=0.2}, draw=none, from=1-1, to=2-2]
	\arrow["{\mathit{post}}"{pos=0.4}, dashed, from=1-2, to=2-2]
	\arrow["{\mathit{adjust}}"', dashed, from=2-1, to=2-2]
\end{tikzcd}
\\[4pt]
(c)
\end{minipage}
\caption{Projection}
\label{fig:projection}
\end{figure}

The \textsf{Project} function takes as inputs two timelines with the same initial state and produces two new timelines continuing from the input timelines and converging on the same final state. It is easier to understand this with diagrams.
In Figure~\ref{fig:projection}(a) the arrow on the left takes the state \mathbox{0} (we elide the obvious type signature) and writes a \mathbox{2} over it. The arrow on the top wraps that \mathbox{0} into a list with one element \mathbox{[\mathbb{1}\is 0]}. The projection function takes these two arrows as input and produces the two dashed arrows such that the diagram \textit{commutes}, meaning the dashed arrow operations are valid on the states at their bases, and they both produce the same state \mathbox{[\mathbb{1}\is 2]}. The way this is achieved is to map the target path of the \textsf{write} operation through the \textsf{ListOf} operation producing the operation \mathbox{\ldotp\mathbb{1} \textsf{ write } 2}. We say that the two \textsf{write} operations have the same \textit{intention}---they ``do the same thing'' in different states---they convert the \mathbox{0} to a \mathbox{2}, even though the \mathbox{0} has moved to a different location. That is the key principle of projection: it produce a right-hand operation that ``does the same thing'' as the left-hand operation despite the top operation having intervened.

Figure~\ref{fig:projection}(b) shows a messier situation. Here the left and top operations are both writing different values to the same location. This is called a \textit{conflict}. The rule is that projection attempts to preserve the intention of the left operation, even if that means overriding the top operation, which as a consequence gets converted to a \textsf{noop} on the bottom.

This last example shows that projection is asymmetric: flipping the diagram on its diagonal to switch left and top yields a different result. And in fact we will have occasion to flip and rotate these diagrams, so we cannot depend on terminology like ``left'' and ``top''. Figure~\ref{fig:projection}(c) shows our orientation-neutral terminology: projection converts the \textit{pre} operation into the \textit{post} operation, preserving its intent after the \textit{base} operation intervenes. The \textit{base} operation is converted into the \textit{adjust} operation preserving as much of its intention as is allowed by the first rule. We also graphically indicate the orientation of the diagram by placing \mathbox{\textsf{P}\!\Rightarrow} into the corner of the initial state.

\begin{figure}[h]
\begin{minipage}[b][][b]{1.6in}
\center
\begin{tikzcd}[column sep=large]
	\bullet & \bullet \\
	\bullet & \bullet
	\arrow["{{p \textsf{ ListOf}}}", from=1-1, to=1-2]
	\arrow["{{p \ldotp\ldotp q  \ A}}"'{pos=0.4}, from=1-1, to=2-1]
	\arrow["{{\textsf{P}\Rightarrow}}"{description, pos=0.2}, draw=none, from=1-1, to=2-2]
	\arrow["{{p \ldotp \mathbb{1} \ldotp\ldotp q \ A}}"{pos=0.4}, dashed, from=1-2, to=2-2]
	\arrow["{{p \textsf{ ListOf}}}"', dashed, from=2-1, to=2-2]
\end{tikzcd}
\\[4pt]
(a)
\end{minipage}
\quad
\begin{minipage}[b][][b]{1.7in}
\center
\begin{tikzcd}[column sep=large]
	\bullet & \bullet \\
	\bullet & \bullet
	\arrow["{{p \textsf{ write } w}}", from=1-1, to=1-2]
	\arrow["{{p \textsf{ write } v}}"'{pos=0.4}, from=1-1, to=2-1]
	\arrow["{{\textsf{P}\Rightarrow}}"{description, pos=0.2}, draw=none, from=1-1, to=2-2]
	\arrow["{{p \textsf{ write } v}}"{pos=0.4}, dashed, from=1-2, to=2-2]
	\arrow["{{p \textsf{ noop}}}"', dashed, from=2-1, to=2-2]
\end{tikzcd}
\\[4pt]
(b)
\end{minipage}
\caption{Projection rules}
\label{fig:projection-rules}
\end{figure}

Projection is defined with rules that can be diagrammed similarly. Figure~\ref{fig:projection-rules}(a) is a rule covering the \textsf{ListOf} example. The target path of the \textsf{ListOf} is abstracted to \mathbox{p}, and the \textsf{write} operation is abstracted into any operation \mathbox{A} targeting \mathbox{p} or deeper, which is indicated with the path concatenation operator in \mathbox{p \ldotp\ldotp q}. Operation \mathbox{A} is projected down into the \textsf{ListOf} with the path \mathbox{p \ldotp \mathbb{1} \ldotp\ldotp q}.

Figure~\ref{fig:projection-rules}(b) is the rule for conflicting writes.

\begin{figure}[h]
\begin{minipage}[b][][b]{1.2in}
\center
\begin{tikzcd}[column sep=large]
	\bullet & \bullet \\
	\bullet & \bullet
	\arrow["{{\mathit{base}}}", from=1-1, to=1-2]
	\arrow["{{\mathit{pre}}}"'{pos=0.4}, dashed, from=1-1, to=2-1]
	\arrow["{{\Leftarrow\textsf{R}}}"{description, pos=0.2}, draw=none, from=1-2, to=2-1]
	\arrow["{{\mathit{post}}}"{pos=0.4}, from=1-2, to=2-2]
	\arrow["{{\mathit{adjust}}}"', dashed, from=2-1, to=2-2]
\end{tikzcd}
\\[4pt]
(a)
\end{minipage}
\quad
\begin{minipage}[b][][b]{1.8in}
\center
\begin{tikzcd}[column sep=large]
	0 & {[\mathbb{1}\is 0]} \\
	2 & {[\mathbb{1}\is 2]}
	\arrow["{{\ldotp \textsf{ ListOf}}}", from=1-1, to=1-2]
	\arrow["{{\ldotp \textsf{ write } 2}}"'{pos=0.4}, dashed, from=1-1, to=2-1]
	\arrow["{{\Leftarrow\textsf{R}}}"{description, pos=0.2}, draw=none, from=1-2, to=2-1]
	\arrow["{{\ldotp\mathbb{1} \textsf{ write } 2}}"{pos=0.4}, from=1-2, to=2-2]
	\arrow["{{\ldotp \textsf{ ListOf}}}"', dashed, from=2-1, to=2-2]
\end{tikzcd}
\\[4pt]
(b)
\end{minipage}
\quad
\begin{minipage}[b][][b]{1.75in}
\center
\begin{tikzcd}
	0 & {[\mathbb{1}\is 0]} \\
	{} & {[\mathbb{1}\is 2\comma e \is \textsf{NaN}]}
	\arrow["{{\ldotp \textsf{ ListOf}}}", from=1-1, to=1-2]
	\arrow["{{\nLeftarrow\textsf{R}}}"{description, pos=0.2}, draw=none, from=1-2, to=2-1]
	\arrow["{{\ldotp \textsf{ append } e}}"{pos=0.4}, from=1-2, to=2-2]
\end{tikzcd}
\\[4pt]
(c)
\end{minipage}
\caption{Retraction}
\label{fig:retraction}
\end{figure}

\subsection{Retraction}

The \textsf{Retract} function goes in the opposite direction of \textsf{Project}. Figure~\ref{fig:retraction}(a) shows that it starts with two timelines \textit{base} and \textit{post} and produces \textit{pre} and \textit{adjust}. We say that \textit{post} is retracted through \textit{base} into \textit{pre}. The goal as with projection is that \textit{pre} preserves the intention of \textit{post} before \textit{base} happened---it ``does the same thing''. Likewise \textit{adjust} preserves the intention of \textit{base} but defers to the first rule.

Figure~\ref{fig:retraction}(b) shows the retraction reversing the projection in Figure~\ref{fig:projection}(a). Often retraction is the inverse of projection as in this case, but not always, and sometimes retraction is not possible at all. In Figure~\ref{fig:retraction}(c) the list has a new element appended to it. There is no way to do the same thing before the list was created. Projection failures indicates a \textit{dependency} between operations. The \textsf{append} operation depends on the prior \textsf{ListOf} operation having created the list and cannot be retracted through it.

Retraction is defined with rules like those for projection so we skip over them in the interest of brevity.
\begin{note}
  Baseline implements projection and retraction rules with procedural if and switch statements, which has become unwieldy as the number of operations has grown, so much so that the implementation is incomplete. It should be possible to define a declarative DSL that exploits the inherent symmetries in the rules: projection is often symmetric in the two input operations; projection and retraction are often inverses. See \S~\ref{discussion}.
\end{note}

\begin{remark}
How do we know if the rules for projection and retraction are correct, that they do indeed preserve the intentions of operations? Our answer is that there is no objective answer to this question. The rules are essentially axioms defining what it means to preserve intention---intuitively doing the same thing. Well-designed rules will agree with intuition and avoid surprises. One of these intuitions is that transfer should be symmetric: transfering an operation across a diff and then transferring it back again should bring back the same operation (or in some cases grounding out into a \textsf{noop}). We call this the \textit{roundtrip law} and we test that our rules obey it.
\end{remark}

\begin{figure}[]
  \vspace{-20pt}
\[
\textsf{Transfer}(a_{1\ldots n}, O, b_{1\ldots m}) = b' \textrm{ in }\quad
\begin{tikzcd}[column sep = huge]
	A_{n-1} & O & B \\
	A_n & {O'} & {B'}
	\arrow["{a_n}"', from=1-1, to=2-1]
	\arrow["{\textsf{R}\Rightarrow}"{description, pos=0.1}, draw=none, from=1-1, to=2-2]
	\arrow["{a_{n-1}\,\ldots\  a_1}"', from=1-2, to=1-1]
	\arrow["{b_1\,\ldots \  b_m}", from=1-2, to=1-3]
	\arrow["o", dashed, from=1-2, to=2-2]
	\arrow["{\textsf{P}\Rightarrow}"{description, pos=0.2}, draw=none, from=1-2, to=2-3]
	\arrow["{b'}", dashed, from=1-3, to=2-3]
	\arrow["{\mathit{adjust}_a}"', dashed, from=2-2, to=2-1]
	\arrow["{\mathit{adjust}_b}", dashed, from=2-2, to=2-3]
\end{tikzcd}\]
\vspace{-16pt}
\caption{Transfer}
\label{fig:transfer}
\end{figure}

\subsection{Across the Multiverse}\label{transfer}

  A \textit{diff} between two states \mathbox{A} and \mathbox{B} is a pair of timelines that branch off some shared state \mathbox{O} resulting in \mathbox{A} and \mathbox{B}, diagrammed like this:
\[\begin{tikzcd}
	A & {} & {} & O & {} & {} & B
	\arrow["{{a_n}}"', from=1-2, to=1-1]
	\arrow[dotted, from=1-3, to=1-2]
	\arrow["{{a_1}}"', from=1-4, to=1-3]
	\arrow["{b_1}", from=1-4, to=1-5]
	\arrow[dotted, from=1-5, to=1-6]
	\arrow["{b_m}", from=1-6, to=1-7]
\end{tikzcd}\]

The two timelines are called the \textit{branches} of the diff and we call their operations the \textit{differences}. Many version control systems use something like a diff by looking for the \textit{last common ancestor} in a tree of branching versions, which is then used for \textit{three-way diffing}~\cite{diff3}. We do not do that---we will discuss how to calculate a diff in \S\ref{diff-synth} but for the moment let us take them as given. The key operation on a diff is to
\textit{transfer} an operation from one side to another, which we define diagrammatically in Figure~\ref{fig:transfer} (where we lift projection and retraction from operations to timelines in the obvious way).

Transfer
retracts \mathbox{a_n} backwards through all earlier operations on the \mathbox{A} timeline onto the original state \mathbox{O} and then projects that operation forwards through all the operations of the \mathbox{B} timeline. The final result is the operation \mathbox{b'} that yields a new version \mathbox{B'}. The effect of the transfer is that the final operation \mathbox{b'} preserves the intention of the original operation \mathbox{a_n} given all the operations traversed while traveling backward in time to \mathbox{O} and then forward to \mathbox{B}. Transfer also computes a new diff between \mathbox{A_n} and \mathbox{B'}: the transferred operation is now shared between them in \mathbox{O'} with branching timelines \mathbox{\mathit{adjust}_a} and \mathbox{\mathit{adjust}_b}. This updated diff is used later for diff optimization (\S\ref{diff-synth}).

In this definition we transferred the last operation of the \mathbox{A} branch---an earlier operation can be transferred by using the corresponding truncation of \mathbox{A}.
Since transfer uses retraction it may fail because of a dependency on an earlier operation in the \mathbox{A} branch. However it is always possible to transfer any operation by first transfering all of its dependencies earlier in the branch. A \textit{sync} action that transfers all operations in the branch will always succeed. We also provide a finer-grained \textit{dependency transfer} function that transfers just the transitive dependencies and so always succeeds.

As an example we extend the TODO refactoring in Figure~\ref{fig:TODO-refactor} into two branches: branch \mathbox{A} reassigns the TODO from Jack to Jill; branch \mathbox{B} converts the assignment into a list, changes the assignment from Jack to Jacque and inserts a new assignment to Tom:
\begin{align*}
  O &= [
    e \is \{ \mathit{what}\is \quotedstring{clean}\comma  \mathit{who}\is \quotedstring{Jack} \}
    ] \\
  a_1 &= \ \ldotp e\ldotp\mathit{who}\textsf{ write } \quotedstring{Jill}\\
  A &= [
    e \is \{ \mathit{what}\is \quotedstring{clean}\comma  \mathit{who}\is \quotedstring{Jill} \}
    ] \\
  b_1 &= \ \ldotp *\ldotp \mathit{who} \textsf{ ListOf } \\
  b_2 &= \ \ldotp e\ldotp\mathit{who}\ldotp\mathbb{1} \textsf{ write } \quotedstring{Jacque} \\
  b_3 &= \ \ldotp e\ldotp\mathit{who} \textsf{ insert } g \textsf{ before } \mathbb{1} \exec \ldotp e\ldotp\mathit{who}\ldotp g \textsf{ write } \quotedstring{Tom} \\
  B &= [
    e \is \{ \mathit{what}\is \quotedstring{clean}\comma  \mathit{who}\is
    [ g \is \quotedstring{Tom} \comma \mathbb{1}\is \quotedstring{Jacque}] \}
    ] \\
  \end{align*}
  \vspace{-30pt}\\
We can transfer these operations individually from one side to the other.
If we transfer \mathbox{b_2} into \mathbox{A} the change from Jack to Jacque gets applied to the \textsf{who} field (showing the change in bold):
\[A' = [e \is \{ \mathit{what}\is \quotedstring{clean}\comma  \mathit{who}\is \textbf{\quotedstring{Jacque}} \}]\]
If we transfer \mathbox{a_1} into \mathbox{B} the change from Jack to Jill gets applied to the list element containing Jack, even though it has been wrapped and shifted:\\
\[B' = [
    e \is \{ \mathit{what}\is \quotedstring{clean}\comma  \mathit{who}\is
    [ g \is \quotedstring{Tom} \comma \mathbb{1}\is \textbf{\quotedstring{Jill}}] \}
]\]


The prior example demonstrates how transference can propagate changes bidirectionally through structural changes. Standard state-based version control techniques would see only a big conflict and require a human to pick out and propagate finer-grained changes correctly.
Operational differencing does this more precisely because it tracks how locations shift through time.

Version control systems like git provide a variety of actions like \textit{merge}, \textit{rebase}, and \textit{cherry-pick} with subtly different effects. We provide transferrance as the sole primitive and then build coarser-grained actions out of sets of transfers. For example all the changes in one branch can be transferred to the other. Just the type changes or just the data changes can be transferred. The dependency transfer action mentioned earlier transfers all transitive dependencies. To \textit{synchronize} two branches all changes are transferred from one to the other and then all changes remaining in the other branch are transferred back. Note that synchronization has a direction: syncing from \mathbox{A} into \mathbox{B} resolves all conflicts in favor of \mathbox{A}.

\begin{casestudy}\label{case-multiplicity}
The preceding example of a TODO refactoring was taken directly from the challenge problem \textbf{Multiplicity Change in Schema}~\cite{challenge-problems}. The problem is that different versions of the TODO application must interoperate after the schema evolution from a single assignee to a list. Specifically, changes to the single value \quotedstring{Jack} should be mapped into the list element copying it and vice-versa, despite insertions in the list. Baseline solves this problem by giving that list element the special ID $\mathbb{1}$ to persistently identify it.
\end{casestudy}

\begin{casestudy}
Case study \ref{case-multiplicity} was extracted from \citeauthor*{Cambria}~\cite[Appendex III]{Cambria}, which additionally asked: what happens when \quotedstring{Jack} is deleted from the list? They want to set the old version's single value to a JavaScript \textsf{null}. Baseline currently handles this situation by refusing to retract the \textsf{delete} backwards through the \textsf{ListOf}, decoupling subsequent changes to the old single value from the new list. Perhaps Baseline should instead map deletion to the empty string and vice-versa (or support proper nullable or optional types analogously). This case study is a good example of how ``doing the right thing'' can depend on the context of use---there is no universal truth.
\end{casestudy}

\subsection{Diff Optimization and Synthesis}\label{diff-synth}

We have been discussing transference assuming we are given a diff that specifies a shared state and timelines branching off of it. The intuition is that the shared state should contain everything the two final states have in common, and the branches should be just the differences. Not all branching timelines will satisfy that intuition. There might be redundant operations duplicated in both branches, either by coincidence or by explicit transfers. There could also be redundant changes within a branch that override earlier ones. All of these sorts of redundancies are undesirable because they add noise to a visualization of differences, and in some cases can cause transfer to produce surprising results. There is a simple algorithm that can be used to produce optimal diffs using transference itself.

Note that in Figure~\ref{fig:transfer} the bottom of the diagram produces a new diff: two new timelines \mathbox{\mathit{adjust}_a} and \mathbox{\mathit{adjust}_b} branching from a new shared state \mathbox{O'}. These \textit{adjusted} timelines eliminate the redundancy of the original operation and its transferred image. Clearly after executing a transfer we should drop the original diff and use the adjusted diff instead. In the same way we can use transfers to remove redundancies in diffs as follows. Given a pair of branching timelines we iterate over all the actions on both sides in order (regardless of interleaving because of the roundtrip law). For each operation we try to transfer it to the other branch. If the transfer succeeds and is a fixpoint (it doesn't alter the state of the other side), then we discard the branches in favor of the adjusted ones and continue the loop. The intuition here is that if the transferred operation makes no difference to the other side then it isn't really a difference, so we append it to the history of the shared state. After we have tried every transfer from both sides the result will be optimal in the sense that the history of the shared state is maximal while the timelines branching off of it are minimal, without redundancies. Note however that this maximal shared state may not have ever actually occurred in either history---it is the \textit{hypothetical} ideal last common ancestor if history had happened in a fortuitously non redundant order. This ideal common ancestor avoids the problems that can be caused by cherry-picking in git~\cite{philomatics-git}.

Diff optimization lets us synthesize an optimal diff from the histories of any two states. A history always starts from the unit value as its initial state. So all histories are branches from the initial state and we just optimize that primordial diff.

\begin{note}
  We left out a detail in the diffing algorithm. Sometimes changes cancel each other out, like an insert followed later by a delete, or an operation followed later by its \textit{undo} (Appendix~\ref{undo}). The diff timelines are simplified by attempting to fully retract each operation---if the retraction results in a \textsf{noop} it was canceled out and the simplified adjusted timeline is used instead.
\end{note}

\subsection{Low-fuss Version Control}

The ability to synthesize diffs for any two histories avoids the need for a \textit{repo} to determine the last common ancestor for operations like merging. As we have seen that can be sub-optimal as a basis for three-way diffing. But more importantly, understanding and maintaining the repo imposes a large mental burden on the users of version control systems. Git is infamous for causing confusion and frustration~\cite{perez13, church2014case, xkcd1597}.

We present the user with a simpler conceptual model: every document has an append-only history. To branch an alternate version of a document, simply copy it. The copies can then diverge, and their histories can be compared at any time to compute their diff, which can then be used to transfer differences between them. We can record copy and transfer events as comments in histories in order to reconstruct a version graph like that maintained in a repo.

This approach is more flexible: since versions and branches no longer belong to a repo they can be shared without sharing the whole repo. They are like files (and likely implemented as files): independent storage units that can participate in overlapping and shifting webs of collaboration. They can be emailed. We regain the flexibility of the ubiquitous folk practice of file copying~\cite{Burnett14, Basman19}, but with the added benefit of being able to diff and merge.

\begin{remark}
  Large projects will still need to maintain a list of all branches, but that could be as lightweight as a file directory. Large projects will also want to compress storage and cache diffs. We believe all these capabilities could be added on top of our approach with far less user-facing complexity than git. But that is not an immediate goal: first we want to optimize for small-scale ad hoc usage.
\end{remark}

\begin{figure}
  \vspace{-8pt}
\begin{subfigure}[b]{35em}\vspace{0pt}
  \sffamily
  \small
  \begin{tabular}{ r|l|r|r|l|l|}
  \multicolumn{2}{l}{orders:}\\
     \hhline{~-----}
     ID & item & quantity & ship\_date & name & address \\
     \hhline{~=====}

     $e_1$ & Anvil & 1 & 2/3/23 & Wile E Coyote & 123 Desert Station \\
     \hhline{~-----}
     $e_2$ & Dynamite & 2 & & Daffy Duck & White Rock Lake \\
     \hhline{~-----}
     $e_3$ & Bird Seed & 1 & & Wile E Coyote & 123 Desert Station \\
     \hhline{~-----}
  \end{tabular}
  \caption{Single orders table}
  \label{fig:tables-single}
\end{subfigure}
\\[-0.5em]~\\
\begin{subfigure}[b]{40em}\vspace{0pt}
  \sffamily
  \small
  \begin{tabular}{ r|l|r|r|r|cr|l|l|}
  \multicolumn{2}{l}{orders:}&\multicolumn{4}{l}{}&\multicolumn{2}{l}{customers:}\\
     \hhline{~----~~--}
     ID & item & quantity & ship\_date & customer && ID & name & address \\
     \hhline{~====~~==}

     $e_1$ & Anvil & 1 & 2/3/23 & $\rightarrow\!e_1$ && $e_1$ & Wile E Coyote & 123 Desert Station \\
     \hhline{~----~~--}
     $e_2$ & Dynamite & 2 & & $\rightarrow\!e_2$ && $e_2$ & Daffy Duck & White Rock Lake \\
     \hhline{~----~~--}
     $e_3$ & Bird Seed & 1 & & $\rightarrow\!e_3$ && $e_3$ & Wile E Coyote & 123 Desert Station \\
     \hhline{~----~~--}
  \end{tabular}
  \caption{Split orders and customers tables}
  \label{fig:tables-split}
\end{subfigure}
\\[-0.5em]~\\
\begin{subfigure}[b]{40em}\vspace{0pt}
  \sffamily
  \small
  \begin{tabular}{ r|l|r|r|r|cr|l|l|}
  \multicolumn{2}{l}{orders:}&\multicolumn{4}{l}{}&\multicolumn{2}{l}{customers:}\\
     \hhline{~----~~--}
     ID & item & quantity & ship\_date & customer && ID & name & address \\
     \hhline{~====~~==}

     $e_1$ & Anvil & 1 & 2/3/23 & $\rightarrow\!e_1$ && $e_1$ & Wile E Coyote & 123 Desert Station \\
     \hhline{~----~~--}
     $e_2$ & Dynamite & 2 & & $\rightarrow\!e_2$ && $e_2$ & Daffy Duck & White Rock Lake \\
     \hhline{~----~~--}
     $e_3$ & Bird Seed & 1 & & $\rightarrow\!e_1$ && \multicolumn{3}{l}{}$e_3$\\
     \hhline{~----~~~~}
  \end{tabular}
  \caption{Deduped customers}
  \label{fig:tables-dedup}
\end{subfigure}
\vspace{-8pt}
\caption{Table schema change}
\label{fig:tables}
\end{figure}

\section{Database Evolution}\label{db-evolution}

In this section we apply Operational Differencing to the problem of \textit{schema evolution} in relational databases. We do this by adding relational database features into our data types and operations.
For example Figure~\ref{fig:tables-single} shows a table of orders for the Acme Corporation. A database professional will immediately see a problem: the customer name and address is duplicated across all orders by the same customer, so any changes to that information would need to be duplicated.
Database textbooks teach to avoid this sort of problem by properly \textit{normalizing} the schema before adding any data. However in practice the need for schema changes often arises after there is data, which must then be \textit{migrated} into the changed schema, typically with ad hoc SQL programs.

One approach a SQL migration program could take is to populate a new \textsf{customers} table from a query projecting out the \textsf{name} and \textsf{address} columns and excluding duplicates. Then it would need to generate primary keys for the deduped customers (the name cannot be the primary key because it is mutable). A new column \textsf{customer} would be added to \textsf{orders} declared as a foreign key containing the customer's primary key. Finally the \textsf{name} and \textsf{address} columns in the \textsf{orders} table would be dropped.

To perform the migration in Baseline we first need to add support for relational data into our types and operations.
We already support tables: they are lists of records, where the elements of the list are the rows and the fields of the record are the columns, which are homogeneously typed as in a relational database.
Throughout the rest of the paper we will refer to any Baseline document containing tables as a \textit{database}.
What is missing so far is \textit{relationships} between tables, which relational databases model through primary and foreign keys. We take a more PL-oriented approach.
Instead of primary keys exposed as user data fields we annotate list elements with IDs that are unique, permanent, and opaque. Instead of foreign keys we add the \textit{link} data type in Figure~\ref{fig:db-addons}: the type of a link defines the table (or list) it ranges over; the value of a link is an element ID in that range, or \mathbox{*} for a null link. We also add the operations defined in Figure~\ref{fig:db-addons}. All operations that move a list must also adjust the path in the type of all links into that list.

\begin{figure}
  \vspace{-8pt}
\figbox{
\[\begin{array}{r@{\ }l|r@{\ }l|r@{\ }l|l}
  \multicolumn{2}{l|}{\textrm{type}} & \multicolumn{2}{l|}{\textrm{value}} & \multicolumn{2}{l|}{\textrm{initial value}}&\\
  \hline

  & \rightarrow\! p & & \rightarrow\! E & & \rightarrow\! *
  & \textrm{link to element $E$ of list $p$ ($*$ is null link)}  \\

\end{array}\]
\vspace{-4pt}
\[\begin{array}{l|l}
  \textrm{Value operations} & \\
  \hline

  p \textsf{ link } E & \textrm{Link to element } E \textsf{ (unlink if $*$)} \\

  \hline

  \textrm{Type operations} & \rule[0pt]{0pt}{16pt}\\
  \hline

  p \textsf{ Link } p' & \textrm{Define $p$ as link to list $p'$} \\

  p \textsf{ Split } p' & \textrm{Split at column $p$ creating linked table $p'$}\\
  p \textsf{ Join } & \textrm{Join link column $p$ deleteing linked table}\\

\end{array}\]
}
\vspace{-4pt}
  \caption{Additional database syntax and operations}
  \label{fig:db-addons}
\end{figure}

The first step in evolving the Acme database is to \textsf{Split} out a \textsf{Customers} table with the following operations:
\[
\begin{array}{ll}
\{ \textsf{orders}\is [ \ldots ]\} \exec & \textrm{(1)}\\
\ldotp \textsf{ Append } \textsf{customers} \exec  & \textrm{(2)}\\
\ldotp \textsf{orders}\ldotp * \ldotp \textsf{name Split customers} \exec & \textrm{(3)}\\
\ldotp \textsf{orders}\ldotp * \ldotp \textsf{name Rename } \quotedstring{customer} & \textrm{(4)}\\
\end{array}\]


(1) Starting from the state in Figure~\ref{fig:tables-single};
(2) The \textsf{Append} operation adds a \textsf{customers} field to hold the new table. (3) The \textsf{Split} operation then splits the \textsf{orders} tables in two starting at the \textsf{name} column, copying those columns into the \textsf{customers} table, the rows of which are given the same IDs as in the \textsf{orders} table.
The original \textsf{name} column in \textsf{orders} is replaced with links to the corresponding \textsf{customers} rows.
(4) The \textsf{Rename} operation renames the links column to be \textsf{customer} (abusing notation by conflating the name and ID) resulting in
Figure~\ref{fig:tables-split}.

The last step of this database evolution is to de-duplicate the customers, for which we use the \textsf{move} operation from Figure~\ref{fig:rich-data}. Thus:
\[
\ldotp \textsf{customers}\ldotp e_3 \textsf{ move } \ldotp \textsf{customers}\ldotp e_1
\]
The \textsf{move} operation merges \mathbox{e_3} into \mathbox{e_1}, resulting in the state shown in  Figure~\ref{fig:tables-dedup}. Note that the order for Bird Seed has had its link to \mathbox{e_3} forwarded to \mathbox{e_1}. This completes the schema evolution.

\begin{note}
The preceding discussion does not reflect the full complexity of our data model. Lists and tables can be nested. Links are set-valued, and can drill into nested tables. We believe this complexity is justified to cleanly unify PL and DB data models. The \textsf{Split} and \textsf{Join} operations are composed from a set of primitive operations on nested tables and links. We have experimented with several alternative sets of primitives, with subtle tradeoffs on transfer semantics, for example how does appending a column get projected through the split operation? See \S~\ref{discussion}.
\end{note}

\begin{casestudy}\label{case-extract}
The preceding example of Acme's database evolution is taken verbatim from the challenge problem \textbf{Extract Entity}~\cite{challenge-problems}. Indeed this problem was a primary motivating example in the design of Baseline.
\end{casestudy}

\begin{casestudy}
\citet{sculpin} study restructuring JSON data. Their example of art exhibition data would be modelled relationally (unnormalized) as a table of exhibits containing nested tables of locations. They map the exhibit title into each location and then extract a single concatenated list of the enhanced locations. Baseline's data model supports nested structures like JSON. The \textsf{Split} operation on tables is actually composed from two more general operations on nested structures. The \textsf{Extract} operation replaces content in possibly nested lists with links to a concatenated list of the contents. The \textsf{Push} operation maps a column of a table through another column containing links. These more general operations can perform the JSON restructuring example.
\end{casestudy}


\subsection{Database Version Control}\label{db-vc}

Schema evolution is typically implemented as a one-time one-way event, which becomes problematic when there are multiple replicas of the database in production. Operational Differencing is more flexible because it can transfer changes forward through the schema evolution and backwards as long as they aren't dependent on the schema evolution itself. This could allow distributed migrations to happen asynchronously.
The scheme evolution operations themselves can be transferred between copies of a database, so migrating a database instance can be done by transferring in the type changes from the dev branch instead of an elaborate DevOps procedure.

For example say that the Order Fulfillment department at Acme Corporation is happy with the old version of the database because they don't have to do a query to find the customer name and address. So they don't want to deploy the schema changes made by the Customer Relationship department. They could instead continue to use the old database and transfer updates to the new database. Likewise when a new order is entered in the new database it can be transferred to Fulfillment in its unnormalized form. But new orders to new customers can't be transferred backwards because customer creation is dependent on the \textsf{Split} operation creating the customers table. In those cases the new order will transfer back with blank name and address fields. This problem is an instance of the classic \textit{View Update Problem}\cite{Bancilhon81, Foster2007}. A better solution for this example uses queries as discussed in \S\ref{formulas}.

Even with the limitations on view updates, Operational Differencing brings many of the advantages of version control to the domain of databases. It is possible to calculate the precise differences between two database instances despite intervening schema changes. It is always possible to transfer all the differences from one to the other, and often possible to transfer individual differences.


\begin{casestudy}\label{case-diverge}
The preceding example is taken from the challenge problem \textbf{Divergence Control for Extract Entity}~\cite{challenge-problems}. Our literature review in that paper found surprisingly little prior work on the problems of deduplication and ID assignment despite that being a commonplace situation in database schema evolution.
It seems managing identity in databases is usually left to the programmer.
Indeed Baseline's solution is not entirely satisfactory either.
We used a \textsf{move} operation to deduplicate the two coyotes \mathbox{e_1} and \mathbox{e_3}. We took that approach because it was the simplest thing that could possibly work, but arguably it doesn't. Say that \mathbox{e_1} changed its address in the old database and we transfer that change forward into the deduped table. Our current implementation drops the change, because \mathbox{e_3} overwrites \mathbox{e_1}. We believe a more intuitive result would be to ``de-deduplicate'' leaving two coyotes with the same name but different addresses.
We plan to do that by offering a proper \textsf{Set} datatype that collapses duplicates into equivalence classes of IDs, retaining ID provenance to make change transfers like the above meet intuitive expectations.
\end{casestudy}

\begin{casestudy}
\citet{herrmann17} study the case of a task management system where they split an \textsf{author} column into a separate table while assigning unique IDs that are used as a foreign key. Their \textsf{Decompose} operation combines our \textsf{Split} operation with deduplication.
Their system is not subject to the deduplication anomalies we discussed above, but at the cost of not supporting divergence of data: multiple schema are maintained as updatable views of a single materialized view.
\end{casestudy}

\section{Operationalized Queries}

We saw in the last section that although the Fulfillment department preferred the unnormalized view of orders it would be forced to migrate to the normalized database to stay fully in sync with all changes. Relational databases offer a better solution: \textit{queries}. In this section we discuss how Operational Differencing can build queries and updatable views. But the simplest solution doesn't even require a query. We can just fork a copy of the database and denormalize it.

The \textsf{Join} operation in Figure~\ref{fig:db-addons}
is the inverse of \textsf{Split}. It replaces a link column with all the columns of the linked table, copying their contents to match the link values.
Specifically:
\[
\ldotp \textsf{orders}\ldotp * \ldotp \textsf{customer Join} \\
\]
resulting in exactly the database we started with in Figure~\ref{fig:tables-single}.
The Fulfillment department can perform this operations directly in the UI, allowing them
to operate in their preferred schema and bidirectionally sync changes with the normalized primary database in the Cutomer Relationship department. Bidirectional sync works now when it didn't before because the \textsf{Join} operation does not create any dependencies. We could make this even more convenient by automatically syncing on all changes. We have essentially built an \textit{updatable view} by doing schema changes on a copy, and we did this in the UI without programming.

This example raises an interesting question: could we generalize it into a general technique for building queries? Instead of writing a query in an abstract language we make a copy of the database and perform schema change and data mutation operations in the UI to \textit{sculpt}~\cite{sculpin} it into the desired form. This sculpted database acts like a query if we automatically sync changes from the primary database into it. Such forward projection is always possible. We can also update the sculpted view and sync those changes back to the real database, limited to those cases where a traditional query-based view would also be updatable.

We call this idea \textit{operationalized queries}. It can be seen as a form of \textit{Programming by Demonstration}~\cite{cypher93-pbd}. We can draw another analogy: query languages are a classic example of pure functional programming that avoids mutation. Pure functions can often be implemented by copying values, mutating them, then returning the result. That is exactly what we are
doing: computing a query by mutating a copy of the database.

\begin{figure}
\vspace{-8pt}
\figbox{
  \[ \begin{array}{r@{\ }l|r@{\ }l|r@{\ }l|l}
  \multicolumn{2}{l|}{\textrm{type}} & \multicolumn{2}{l|}{\textrm{value}} & \multicolumn{2}{l|}{\textrm{initial value}}&\\
  \hline

  & \langle \mathit{op}\fatsemi \ldots \ \uparrow\! p \rangle &&
  && &
  \textrm{Formula returning value at $p$.} \\

\end{array}\]
\vspace{-4pt}
\[
\begin{array}{l|l}
  \textrm{Value operations} & \\
  \hline

  p \textsf{ deletePresent } E & \textrm{Delete rows where column $p$ has non-initial value.} \\
  p \textsf{ deleteAbsent } E & \textrm{Delete rows where column $p$ has initial value.} \\


  \hline

  \textrm{Type operations} & \rule[0pt]{0pt}{16pt}\\
  \hline

\end{array}\]
}
\vspace{-4pt}
  \caption{Additional query syntax and operations}
  \label{fig:query-addons}
\end{figure}

It takes more than a join operation to make a query language. We can take another step with a simple selection predicate. When the Fulfillment department ships an order it sets the date in the \textsf{ship\_date} field. They would like a view showing just the pending orders, which lack a ship date. We can further sculpt the view with the \textsf{deletePresent} operation in Figure~\ref{fig:query-addons}:
\[
\ldotp \textsf{orders}\ldotp * \ldotp \textsf{ship\_date deletePresent} \\
\]
The target path is the type of the \textsf{ship\_date} column---all orders with a non-blank value in that column are deleted, producing the desired view. When a ship date is entered into this view the change will transfer back into the primary database, which has the effect of dropping that order from the recalculated view.

\subsection{Formulas as Future Timelines}\label{formulas}

We have seen how an auto-synced branch of a database can serve as a query on it, and how we can build that query by sculpting the database in the UI. But that leaves us having to hop around different branches to view queries. It is more convenient to embed multiple queries into the database itself as \textit{formulas}. A field defined as a formula has its value computed by evaluating the formula, like a formula in a spreadsheet cell. Matching the form of operational queries, we define a formula as a type in Figure~\ref{fig:query-addons} consisting of a timeline of operations ending with a \textit{return path}. A formula is evaluated by executing the timeline of operations starting with the current state of the database and then returning the final value of the return path. In this sense a formula is a \textit{future timeline}---it is a speculatively executed branch off of the present that returns a value from that hypothetical future.

Formulas as future timelines makes it easy to use Programming by Demonstration~\cite{cypher93-pbd}. In the previous example we created a query by making a copy of the database and ``sculpting'' it into the result we wanted with \textsf{Join} and \textsf{deletePresent} operations. The UI lets us view the differences between the original and sculpted databases (Appendix \ref{UI}) and provides a command \textit{Transfer into Formula} that takes the timeline of differences on one branch of the diff and injects them as a formula into the other side. The return path of the formula is set from the location currently selected in the UI. Applied to the previous example after selecting the denormalized \textsf{orders} table it will append a formula to the normalized database with the operations:
\[
\begin{array}{l}
  \ldotp \textsf{ Append } \textit{new-formula} \exec\\
  \textit{new-formula} \textsf{ Define } \\
  \quad \langle
\ \ldotp \textsf{orders}\ldotp * \ldotp \textsf{customer Join} \exec
\ldotp \textsf{orders}\ldotp * \ldotp \textsf{ship\_date deletePresent} \exec
\!\!\uparrow \ \ldotp \textsf{orders}\
  \rangle\\
\end{array}
\]
Henceforth the \textit{new-formula} field's value will be the result of a query  denormalizing and filtering the current contents of the orders table.




\paragraph{Query Rewriting}

There is another benefit to treating queries as timelines of operations: operation transfer gives us \textit{query rewriting}~\cite{curino08, herrmann17} for free. When a database schema changes, some of the queries written for it may need to change too. This scenario is called forward rewriting---backward rewriting is backporting a query on a new schema to an older one.
In practice query rewriting is done manually by SQL programmers. The just-cited research adapts query optimization techniques to build specialized rewriting algorithms.
Baseline handles query rewriting with the existing machinery of transferrance: the \textsf{Define} operation defining the query is transferred through the schema change operations.

For example, we can transfer the prior example query on the normalized database back into the original unnormalized database we started with in Figure~\ref{fig:tables-single}. Doing so retracts the \textsf{Define} operation backwards through the \textsf{Split} operation that normalized the database (ignoring the deduping \textsf{move} because it is a value operation that doesn't affect the schema). Retracting a formula definition simply retracts its timeline of operations. Thus the formula's \textsf{Join} operation retracts through the database branch's \textsf{Split} operation, which cancels out into a \textsf{noop}, which is dropped from the formula. The formula's \textsf{deletePresent} operations retracts without change. The result of the transfer is executed on the original database:
\[
\begin{array}{l}
  \ldotp \textsf{ Append } \textit{new-formula} \exec\\
  \ldotp \textit{new-formula} \textsf{ Define } \\
  \quad \langle
\ \ldotp \textsf{orders}\ldotp * \ldotp \textsf{ship\_date deletePresent} \exec
\!\!\uparrow \ \ldotp \textsf{orders}\
  \rangle\\
\end{array}
\]
which defines a query to filter the pending orders from the unnormalized orders table.

Forward rewriting is more complicated. It occurs when a formula definition is transferred into a branch containing type differences (i.e., schema changes). Forward rewriting is also done on all formulas within a database whenever a type operation is executed. To transfer a formula forward through a type operation its \textit{inverse} (Appendix~\ref{undo}) is prepended to the formula. The intuition here is that the query was written to apply to the old schema, so to preserve its meaning in the new schema we must first convert the schema back again.
For example if we were to transfer the prior backported query forward again through the \textsf{Split} operation then the inverse operation \textsf{Join} would be prepended to the \textsf{deletePresent} operation, correctly reconstructing the original query.

\begin{casestudy}\label{case-query}
This example of query rewriting is a partial solution to the challenge problem \textbf{Code Co-evolution for Extract Entity}~\cite{challenge-problems}. A complete solution would first require a fully general query language. In particular, selection was handled as a special case of testing for blank values; in general a language of predicates is needed.
We conjecture that just as queries can be seen as timelines of operations, so can pure functional expressions, as a series of imperative operations on a result value. The hope is that by consistently using timelines as the mechanism of language abstraction the semantics of operational transfer through these timelines would provide refactoring and rewriting ``for free''.
\end{casestudy}

\begin{casestudy}
\citet{curino08} first introduced \textit{Schema Modification Operations} using the example of a complex Wikipedia schema evolution. It used selection operations with predicates as discussed above. It also involved making a temporary copy of a table and unioning with another table. Baseline currently does not have either of those operations. They appear straightforward with the caveat that changes to a copy can't be retracted into the original without breaking the roundtrip law.
\end{casestudy}

\section{Evaluation \& Discussion}\label{discussion}

To evaluate the capabilities of Baseline we have examined case studies taken from the research literature. These include the challenge problems identified in our prior paper
\textit{Schema Evolution in Interactive Programming Systems}~\cite{challenge-problems}. We have already discussed four of these challenge problems in case studies \ref{case-multiplicity}, \ref{case-extract}, \ref{case-diverge}, and \ref{case-query}. The other four challenge problems reveal limitations, as follows.

\begin{casestudy}
The challenge problem \textbf{Structured Document Edits}~\cite{challenge-problems} is to
take comma-separated strings in a table column and split them into multiple columns. Changes should be transferable bidirectionally between the new document and a variant of the old. We have implemented a solution using a \textsf{Match} operation that takes a regular expression. This operation takes a string column and extracts the first match group of the regular expression into another column. While this solves the problem as stated we are not satisfied because the use of regular expressions violates Baseline's principle that operations correspond to concrete direct manipulations in the UI. We hope that the language of predicates conjectured in case~\ref{case-query} generalizes into a parsing language while still allowing Programming by Demonstration.
\end{casestudy}

\begin{casestudy}
The challenge problems \textbf{Live State Type Evolution}, \textbf{Code Co-Evolution for Structured Document Edits} and
\textbf{Live Editing Domain-Specific Languages}~\cite{challenge-problems} all require programming language capabilities not yet supported in Baseline.
\end{casestudy}

Baseline is foremost an interactive programming system.
We evaluate and situate its design in Appendix~\ref{tech-dims} using the \textit{Technical Dimensions of Programming Systems}~\cite{techdims} framework.
Much of the work on Baseline has been on its user interface, and especially how to visualize structural differences. Unfortunately space limitations have prevented us from discussing this aspect, which is summarized in Appendix~\ref{UI}.
Ultimately Baseline must be judged in terms of usability.
The playable demo described in Appendix~\ref{demo} has provided some qualitative feedback and could become the basis of a controlled experiment.

There are two main conclusions from our evaluations. First, Baseline offers unique capabilities for version control on databases and documents, the need for which has been documented in field studies~\cite{Burnett14, Basman19}.
Operational Differencing can ``cherry pick'' and merge changes without conflict across substantial schema changes. Doing so by direct manipulation without learning git and without laborious conflict resolution is a substantial benefit to non-technical users.

The second, unavoidable, conclusion is that Baseline's operations and transfer rules are worryingly complex and still incomplete. We are frankly disappointed that an elegant algebra like that of relational databases has not emerged yet. Nevertheless we have a strong intuition that there is a simpler more elegant theory struggling to get out. We report on our interim results here in the hope of motivating future research taking new angles of attack. One angle we want to explore is generalizing the data model into a graph to distill a smaller and complete algebra of primitive operations.

Specific technical problems have been discussed where they arose in the case studies, perhaps most notably the need to support a proper set datatype. An overall problem is that implementing transfer rules in procedural code has proven to be unsustainable. We propose to define a logic programming DSL for specifying these rules more succinctly by exploiting their symmetries: \textsf{Project} and  \textsf{Retract} are often inverses of each other and their rules are often transposeable.






\section{Related Work}

\myparagraph{OT and CRDT}

Operational Differencing is related to
Operational Transformation (OT)~\cite{ceda-ot, Ellis89, Ressel96, Oster06} and Convergent Replicated DataTypes (CRDTs)~\cite{Shapiro11}.
Those techniques ensure that multiple replicas distributed across a network will converge to the same state regardless of the order in which operations propagate across the network. To ensure convergence they predetermine how conflicts will be resolved when the operations are first executed. However in version control we need to let humans decide how conflicts will be resolved long after the fact. Version control is about managing divergence not just convergence. On the other hand, most OT and all CRDT algorithms are decentralized, whereas extending our approach to do replication would require a centralized \textit{primary} whose order of operations decides conflict resolution for everyone consistently.

While we have different requirements it would be fair to say that Operational Differencing is a generalization of OT, and indeed was inspired by it. Our Project and Retract functions can be seen as generalizing OT's Inclusion Transformation (IT) and Exclusion Transformation (ET). The difference is that our functions return a pair of operations not just one, retraction may fail, and they do not satisfy the correctness properties TP1 and TP2. In a sense we generalize (weaken) OT by removing predetermination of conflict resolution and insertion ordering (making projection non-commutative) and allowing retraction to fail because of dependencies (making it partial). This weakening makes our projection and retraction rules significantly more complex.

There is work on adding branching and conflict resolution to CRDTs~\cite{patchwork, automerge-conflicts} and to DB replication~\cite{couchdb-conflicts}.
CEDA~\cite{ceda} is an OT-based multi-paradigm distributed database.
Synql~\cite{ignat-synql} is a CRDT-based approach to replicating relational databases respecting integrity constraints. Braid~\cite{braid} is an attempt to unify OT and CRDT.

The power of CRDTs is their monotonic semantics but that is also their weakness. They can't go backwards, as in OT's ET and our \textsf{Retract}. Cherry picking a change requires moving it backwards to the most recent ancestor. Going backwards is also essential for undo, especially selective undo (Appendix~\ref{undo}).

\myparagraph{Databases}

There is a vast literature on schema evolution in databases, however we will limit the discussion here to the minority of operation-based approaches. A fuller survey can be found in \citet{challenge-problems}.

EvolveDB\cite{evolvedb} reverse-engineers the schema into a richer data model and then tracks operations on that model within an IDE. This operation history is used to generate SQL scripts to evolve the database. EdgeDB~\cite{edgedb} resolves the ambiguities of state-based schema comparison by asking the developer to clarify what operations were intended, with some answers able to specify custom migration code in a proprietary query language.

\citet{curino08} initiated a cascade of research on Database Evolution Languages (DELs) using Schema Modification Operators (SMOs) which take
 ``as input a relational schema and the underlying database, and produces as output a (modified) version of the input schema and a migrated version of the database''. SMOs can also rewrite queries to accommodate schema changes.
 \citet{herrmann15} defined a relationally complete DEL and then extended it into a Bidirectional Database Evolution Language (BiDEL)~\cite{herrmann17} that has bidirectional transformation capabilities comparable to Baseline, although limited to supporting concurrent schema within one database.

Schema evolution is also a problem for NoSQL~\cite{sadalage12} databases. While such databases are sometimes called ``schemaless'' in effect that means the schema is left implicit and tools must try to infer it~\cite{storl20, storl22}. \citet{Cambria} uses lenses~\cite{Foster2007} for bidirectional transformation of JSON. \citet{scherzinger13} define a set of operators like the relational SMOs discussed above. \citeauthor*{chillon22}~\cite{chillon22, chillon23} offer a comprehensive set of SMOs over a data model unifying SQL and NoSQL that can do query rewriting.

Schema evolution has also been studied for Object Oriented Databases(OODB)~\cite{li99, banerjee87}. Smalltalk~\cite{Goldberg80} is itself an OODB, persisting all object instances in an \textit{image file} with some evolution capabilities incorporated in the programming environment~\cite[pp.252-272]{Goldberg80}. Gemstone turns the Smalltalk image into a production-quality database and accordingly provides a complex schema evolution API~\cite{Gemstone}. \citet{kamina25} extend a relational DEL for persistent objects.

\myparagraph{Data Science}
AI~assistants~\cite{AIassistants} provide a mechanism for interactively correcting
transformations inferred by state-based schema comparisons.
Wrangler~\cite{kandel11} observes interactive manipulation of sampled data to write data transformation programs.
\citet{petersohn20} present an algebra of operations on Python dataframes that include shape transformations.

\myparagraph{Bidirectional Transformations}
 Bidirectional Transformations~\cite{czarnecki2009bidirectional, hu2024bidirectional} have been a rich field of research.
Coupled Transformations~\cite{Berdaguer07, alcino06, Cleve2006} express transformations coordinated between types and their instances by encoding them into functions on Haskell GADTs built with strategy combinators. \citet{JVisser08} extends the encoding to transform queries written in point-free style. \citet{lammel16} recapitulates coupled transformations within logic programming which extends it to transform logic programs.

Lenses~\cite{Foster2007,edit-lenses} are a linguistic approach to bidirectional transformations that have been the subject of much research.
 ~\citet{carvalho24} use lenses explicitly specified inside code to express its evolution. Lenses are an elegant construct but they are essentially first-order: they can transform data changes bidirectionally through schema changes, but they cannot transform schema changes through other schema changes. That is required in order to merge branches whose schema have diverged. A key benefit of Baseline is treating schema and data changes uniformly.

\myparagraph{Model Driven Engineering}
Unlike the textual artifacts involved in other domains, models typically assign unique identifiers enabling more precise differencing and mergeing~\cite{alanen2003}.
Models are often themselves modeled by a metamodel, which lifts the evolution problem to the metalevel: models must be migrated through the evolution of their metamodel, analogously to changing the grammar of a programming language.~\cite{Herrmannsdoerfer11}. \citet{Cicchetti11} study metamodel evolution in the context of divergent changes.
\citet{vermolen11} define a DSL for evolving a data model by generating SQL migrations on the backing database. They develop comparable capabilities to Baseline, although non-interactive. They consider the evolution of OO-style subclass hierarchies which goes beyond our and most other work.
\citet{SemanticDeltas} proposes a research agenda for thoroughly live DSL programming focusing on \emph{Semantic Deltas} that unify edit-time and run-time change.
\citet{exelmans25} offer an operation-based solution to our challenge problem of \textit{Live Editing Domain-Specific Languages} with connections to our approach.

\section{Conclusion}

Baseline unifies data management across multiple dimensions: mutation over time, collaboration across space, and evolution through design changes. It leverages \textit{Operational Differencing}, a new technique for managing data in terms of high-level operations that include refactorings and schema changes.
Integrating schema change operations into a rich data model makes schema evolution a first-class concern instead of an afterthought.
Operational Differencing offers a new kind of version control on data that is finer-grained and higher-fidelity than conventional state-based approaches and is able to work across schema evolutions. It also offers a simplified conceptual model for ad hoc usage.
Our preliminary exploration of an operationalized query language opens an avenue towards incorporating programming language capabilities.

We have identified significant limitations requiring further research, but we
believe that Baseline demonstrates, at least, that operation-based approaches to data management have been under-explored and promise valuable new capabilities.

\acks
We thank Joshua Horowitz, Devamardeep Hayatpur, and the anonymous reviewers for their helpful comments.

\appendix

\section{Deletion and Tombstones}\label{deletion}

\citet{Oster06} first reported a flaw in how all prior OT algorithms handled deletion. Insertions on either side of a deletion can become conflated and misordered.
The now-standard solution is that deletions should leave behind a \textit{tombstone} as a placeholder to keep the insertions on either side separate. The tombstone is hidden from the user, but still occupies space in the string.

The current implementation of Baseline does not leave tombstones in the state. Instead they are tracked in the affected \textsf{insert} operations. Each insert operation includes a sequence of IDs of ``virtual tombstones'' that record the fact that the insertion was shifted right over a deletion.
When an insert operation is projected through a deletion of its insertion point (the \textsf{before} paramater) the insertion point is shifted to the right of the deletion and the ID of the deleted element is appended to the list of tombstones in the insert operation. When inserts at the same point are projected through each other their tombstone lists are compared to decide which to shift in front of the other. We believe that this is a novel solution to the tombstone problem,
however it substantially complicates the projection and retraction rules for inserts and deletes.

State tombstones are easier to understand, so we have adopted them in this paper's description of Baseline. Tombstones have also proven useful in the UI as placeholders when diffing insertions and deletions. We plan to adopt them in the next implementation.

\section{Selective Undo}\label{undo}

\textit{Undo} is an essential feature when editing any sort of information, although it is frustrating that the common approach only allows undoing the last operation. Often we only realize mistakes later. \textit{Selective undo}~\cite{berlage94} allows the user to undo past operations. Intuitively, undoing a past operation should result in what ``would have happened'' if that operation had never occurred but all subsequent operations still did.

It is tempting to think of selective undo as simply deleting an operation from the timeline. There are two problems with this view. Firstly, subsequent operations might depend upon the effects of that operation. Subsequent operations have to somehow be scrubbed of all traces of the undone operation. Secondly, in the setting of collaborative replicas or version control we can't just change history---we need to calculate the delta to the current state that will result in the same thing as that hypothetical changed history. In this way the undo operation can be shared with collaborators like any other operation.

Selective undo has been implemented in OT~\cite{sun00, weiss08} and Baseline takes a similar approach, diagrammed below.
\[\begin{tikzcd}[column sep=large]
	{\{\}} & {A_{i-1}} & {A_i} & {A_n} \\
	&& {A_{i-1}} & {A'}
	\arrow["{{{a_{1 \cdots i-1}}}}", from=1-1, to=1-2]
	\arrow["{{{a_i}}}", from=1-2, to=1-3]
	\arrow["{{{a_{i+1 \cdots n}}}}", from=1-3, to=1-4]
	\arrow["{{{- \langle A_{i-1}\fatsemi\  a_i \rangle}}}"'{pos=0.4}, from=1-3, to=2-3]
	\arrow["{{\begin{array}{c} \textsf{P}\Downarrow \end{array}}}"{description, pos=0.2}, draw=none, from=1-3, to=2-4]
	\arrow["{{{\Delta}}}"{pos=0.4}, dashed, from=1-4, to=2-4]
	\arrow[dashed, from=2-3, to=2-4]
\end{tikzcd}\]

The top row of the diagram shows the complete history of document \mathbox{A} starting at the unit value \mathbox{\{\}} and ending with state \mathbox{A_n}. To undo a past operation \mathbox{a_i} we first calculate its \textit{inverse} \mathbox{- \langle A_{i-1}\!\fatsemi a_i \rangle} that takes the state \mathbox{A_i} back to the state \mathbox{A_{i-1}}. In general the inverse takes a timeline and produces a sequence of operations that may depend upon the initial state. For example if \mathbox{a_i} is a \textsf{delete} operation its inverse will first re-insert the deleted element at the proper location and then completely restore its value in the prior state.

Given the inverse operations we then project the subsequent timeline through them. Note the downward direction of the projection, which causes subsequent operations to override the inverse. For example if we undo a past \textsf{write} operation but there were subsequent writes to the same location then undoing should have no effect. The projection produces the \mathbox{\Delta} operations on the right hand side, which is the delta we are looking for. We execute the delta on \mathbox{A} producing the result of the undo in \mathbox{A'}. The projection also produces the dashed arrow on the bottom, which is not utilized, but is of interest because it is that hypothetical history scrubbed of the effects of the undone operation that we imagined earlier.

\section{Visualizing Structure and Change}\label{UI}

Usability is a primary goal of Baseline. We believe that people should have the same power as programs, entailing that all state is visualized in a UI and that all operations are manifested as direct manipulations in it. The same goes for our version control capabilities: differences are visualized and can be transferred. There is a playable demo (see Appendix~\ref{demo}) and a short video presentation at  \url{https://www.hytradboi.com/2025/3b6de0f0-c61c-4e70-9bae-cca5a0e5bb7b-db-usability-as-if}.

\begin{figure}[h]
\includegraphics[width=\textwidth]{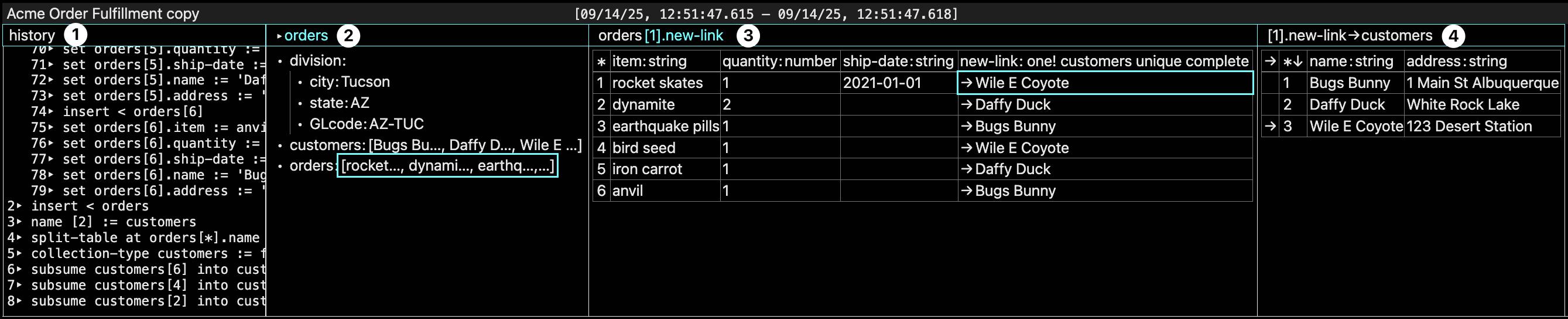}
\caption{A selection is a path through Miller columns containing indented outlines and flat tables.}
\label{fig:UI}
\end{figure}

The following discussion concentrates on the rationale behind the key UI design decisions.
Figure~\ref{fig:UI} displays a screen shot. \circledtext{1} The left hand column shows the history of operations on this database. History sidebars are a common technique but we were particularly inspired by Wrangler~\cite{kandel11}.

The three columns to the right are Miller columns~\cite{miller-columns} that display a path drilling into and across nested data structures. Usually Miller columns display a simple list, as in the MacOS Finder column view of directories. We elaborate that convention to show in each column either an indented outline or a flat table. Nested structures are indicated with summaries that when clicked open a new column to the right. Cyan outlines show the selected path through the columns. \circledtext{2} is the top-level structure of this document. \circledtext{3} is the \textsf{orders} table within it. The user has selected a link to the \textsf{customers} table, which is opened in the column \circledtext{4}. Selection not only drills into nested structures but also follows links crossing the hierarchy, as if the target table of the link was nested inside it. This design was inspired as a reaction to Ultorg~\cite{bakke:phdthesis, ultorg} which uses nested tables to represent queries crossing a flat relational database. Ultorg provides a single view of a nested structure whereas our use of Miller columns provides only a view of the context of a path within the structure. The compensation for that restriction is that each column contains a visually simpler and context-independent layout, keeping neighbors at each level adjacent to each other rather than being splayed across lower layers.

\begin{figure}[h]
\includegraphics[width=\textwidth]{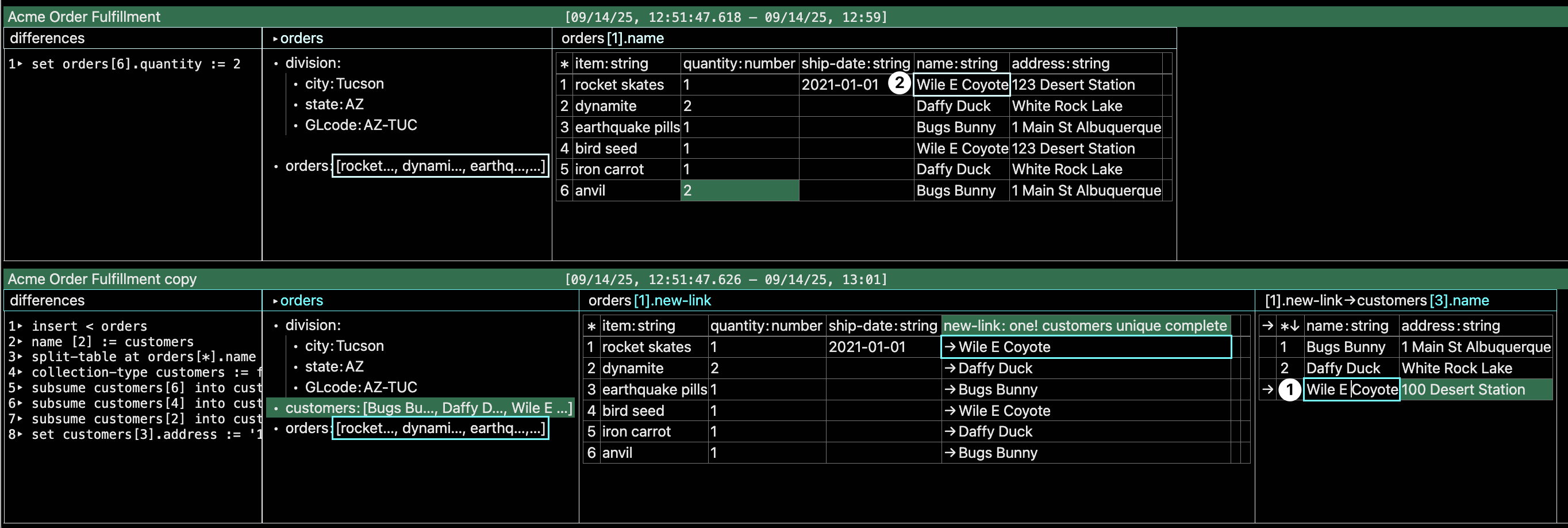}
\caption{A diff is displayed across a split screen with coordinated selection.}
\label{fig:Diff}
\end{figure}

Figure~\ref{fig:Diff} displays our representation of diffs. The challenge is that Operational Differencing can do fine-grained comparisons across schema changes. We split the screen in half vertically to compare two databases. Here we are displaying the differences due to database normalization between Figure~\ref{fig:tables-single} in the top half and Figure~\ref{fig:tables-dedup} in the bottom half. Differences on each side are given a green background, and the context menu on them provides a command to transfer that difference to the other side.

The hard problem is not showing what has \textit{changed} but showing what is the \textit{same} across complex structural transformations.
We use \textit{coordinated selection}:  At \circledtext{1} the cyan outline indicates we have selected the \textsf{customers} table cell containing \textsf{Wile E Coyote}. At \circledtext{2} the corresponding cell in the other database has been outlined in white, even though it is in the \textsf{orders} table. We automatically compute the corresponding location of the selection by transfering a \textsf{noop} operation to the other side.

We first tried conventional representations of textual diffs, both side-by-side and interleaved (AKA unified). They have the advantage of keeping corresponding locations close to each other on screen, but only when the differences are limited to inserting and deleting text. They did not adapt well to complex data structures and complex structural transformations. Our split-screen approach handles these cases as well, and has the advantage that the diff view keeps the layout of the data the same as in a normal view.

\section{The Playable Demo}\label{demo}

It is difficult to convey the experience of an interactive user interface in a written narrative, so we have developed a \textit{Playable Demo}. We adopt \textit{onboarding} techniques that commercial software uses to teach new users. A scaffold over the UI walks the user through a demo scenario, stepping through a written narrative that prompts each user action. Users that follow the main quest are given a frustration-free experience going at their own pace. The more adventurous are free to roam at their own risk, but can always respawn where they left the path. The demo can be accessed at \url{https://thebaseline.dev/Prog26submission/}. We encourage viewers to share their experience in a survey at the end.

Can a playable demo do a better job explaining new user interface ideas than
written narratives and recorded demos? Our early results are mixed.

\begin{itemize}
\item Using an open source onboarding library was a mistake. A fully integrated experience requires a fully custom implementation.
\item To make a playable demo robust it needs to know exactly what each user step should do and validate it. Specifying that with code is laborious (so much so that we did not do it). Instead we should use the history recorded in a play-through as an oracle. That would also allow ``fast forwarding'' through the demo, a repeated user request.
\item A well-presented demo can lull viewers into suspending disbelief. A playable demo is brutally realistic. Lack of polish is glaringly obvious. Going off script quickly reveals holes in the prototype implementation. These effects make playable demos a risky proposition in competitive academic venues.
\item Despite these negatives, users consistently report that when following the script they feel they understand what is being presented.

\end{itemize}

\section{Technical Dimensions}\label{tech-dims}
We can heuristically evaluate Baseline using the \textit{Technical Dimensions of Programming Systems}~\cite{techdims} framework.

\myparagraph{Interaction} Strictly speaking Baseline has two \textbf{modes of interaction}: user and developer, however user mode is a strict subset of developer mode (no type operations) so it is fair to say there is really only one mode of interaction.
Operationalized queries minimize \textbf{abstraction construction} since they correspond exactly to interactive actions in the UI, suporting Programming by Demonstration.
Operationalizing all change shortens \textbf{feedback loops}: The Gulf of Evaluation is reduced because all operations are concrete direct manipulations; The Gulf of Interpretation is reduced by visualizing changes in terms of concrete difference operations. Version control of data enables collaborative feedback loops that are not practical with conventional techniques.

\myparagraph{Notation} Baseline currently has a \textbf{primary notation} that is both a \textbf{surface and internal notation}: sequences of operations. However we expect general programming language features will use a superset \textbf{overlapping notation} that evaluates into the base one. Arguably because Baseline involves a rich set of operations they do not form a \textbf{uniform notation} and generate a complex \textbf{expression geography}.

\myparagraph{Conceptual structure} Baseline is attempting to build a common mechanism for all dimensions of change management, in order to maximize \textbf{conceptual integrity} even at the cost of \textbf{openness}. Our approach to \textbf{commonality} rejects the  dichotomy between static/explicit and dynamic/implicit structure: all structure is explicit but can be dynamically refactored as needed. \textbf{Composability} is uniformly through operation sequencing.

\myparagraph{Customizability} Operations are executed interactively without \textbf{staging}, but operations cen be staged in branches and transferred later when desired. Baseline is not \textbf{self-sustaining} because it is not implemented in itself and we do not hold that as a goal. Our ID-annotated tree structures provide some level of \textbf{addressing and externalizability}.

\myparagraph{Complexity, Errors, and Adoptability} Baseline is still very immature in these dimensions.

\printbibliography

@inproceedings{alanen2003,
  author    = {Marcus Alanen and
               Ivan Porres},
  editor    = {Perdita Stevens and
               Jon Whittle and
               Grady Booch},
  title     = {Difference and Union of Models},
  booktitle = {{\guillemotleft}UML{\guillemotright} 2003 - The Unified Modeling Language,
               Modeling Languages and Applications, 6th International Conference,
               San Francisco, CA, USA, October 20-24, 2003, Proceedings},
  series    = {Lecture Notes in Computer Science},
  volume    = {2863},
  pages     = {2--17},
  publisher = {Springer},
  year      = {2003},
  doi       = {10.1007/978-3-540-45221-8\_2},
}

@inproceedings{alcino06,
  author    = {Cunha, Alcino
               and Oliveira, Jos{\'e} Nuno
               and Visser, Joost},
  editor    = {Misra, Jayadev
               and Nipkow, Tobias
               and Sekerinski, Emil},
  title     = {Type-Safe Two-Level Data Transformation},
  booktitle = {FM 2006: Formal Methods},
  year      = {2006},
  publisher = {Springer Berlin Heidelberg},
  address   = {Berlin, Heidelberg},
  pages     = {284--299},
  isbn      = {978-3-540-37216-5}
}

@article{Bancilhon81,
  author     = {Bancilhon, F. and Spyratos, N.},
  title      = {Update semantics of relational views},
  year       = {1981},
  issue_date = {Dec. 1981},
  publisher  = {Association for Computing Machinery},
  address    = {New York, NY, USA},
  volume     = {6},
  number     = {4},
  issn       = {0362-5915},
  doi        = {10.1145/319628.319634},
  journal    = {ACM Transactions on Database Systems},
  month      = dec,
  pages      = {557–575},
  numpages   = {19}
}

@inproceedings{banerjee87,
  author    = {Banerjee, Jay and Kim, Won and Kim, Hyoung-Joo and Korth, Henry F.},
  title     = {Semantics and implementation of schema evolution in object-oriented databases},
  year      = {1987},
  isbn      = {0897912365},
  publisher = {Association for Computing Machinery},
  doi       = {10.1145/38713.38748},
  booktitle = {Proceedings of the 1987 ACM SIGMOD International Conference on Management of Data},
  pages     = {311--322},
  numpages  = {12},
  location  = {San Francisco, California, USA},
  series    = {SIGMOD '87}
}

@inproceedings{Basman19,
  author    = {Antranig Basman},
  editor    = {Mariana Marasoiu and Luke Church and Lindsay Marshall},
  title     = {The Naturalist's Friend - {A} case study and blueprint for pluralist data tools and infrastructure},
  booktitle = {Proceedings of the 30th Annual Workshop of the Psychology of Programming Interest Group, {PPIG} 2019, Newcastle University, UK, August 28 - 30, 2019},
  publisher = {Psychology of Programming Interest Group},
  month     = aug,
  year      = {2019},
  url       = {https://ppig.org/papers/2019-ppig-30th-basman/},
}

@inproceedings{Berdaguer07,
  author    = {Berdaguer, Pablo and Cunha, Alcino and Pacheco, Hugo and Visser, Joost},
  title     = {Coupled schema transformation and data conversion for XML and SQL},
  year      = {2007},
  isbn      = {3540696083},
  publisher = {Springer-Verlag},
  address   = {Berlin, Heidelberg},
  doi       = {10.1007/978-3-540-69611-7_19},
  booktitle = {Proceedings of the 9th International Conference on Practical Aspects of Declarative Languages},
  pages     = {290–304},
  numpages  = {15},
  location  = {Nice, France},
  series    = {PADL'07}
}

@inproceedings{Burnett14,
  author    = {Burnett, Margaret M. and Myers, Brad A.},
  title     = {Future of End-User Software Engineering: Beyond the Silos},
  year      = {2014},
  isbn      = {9781450328654},
  publisher = {Association for Computing Machinery},
  doi       = {10.1145/2593882.2593896},
  booktitle = {Future of Software Engineering Proceedings},
  pages     = {201--211},
  numpages  = {11},
  location  = {Hyderabad, India},
  series    = {FOSE 2014}
}

@online{Cambria,
title={Project Cambria: Translate your data with lenses},
author={Geoffrey Litt and Peter van Hardenberg and Henry Orion},
url={https://www.inkandswitch.com/cambria.html},
year={2020},
urldate={2020-10-01}
}

@misc{chillon22,
  title         = {A Taxonomy of Schema Changes for NoSQL Databases},
  author        = {Alberto Hern{\'a}ndez Chill{\'o}n and Meike Klettke and Diego Sevilla Ruiz and Jes{\'u}s Garc{\'i}a Molina},
  year          = {2022},
  eprint        = {2205.11660},
  archiveprefix = {arXiv},
  primaryclass  = {cs.DB}
}

@inproceedings{chillon23 ,
  title     = {Propagating Schema Changes to Code: An Approach Based on a Unified Data Model},
  volume    = {3379},
  issn      = {1613-0073},
  booktitle = {CEUR Workshop Proceedings},
  publisher = {CEUR-WS},
  author    = {Chillón, Alberto Hernández and Molina, Jesús García and Hoyos, José Ramón and Ortín, María José},
  year      = {2023}
}

@inproceedings{Cicchetti11,
  author    = {Cicchetti, Antonio and Ciccozzi, Federico and Leveque, Thomas and Pierantonio, Alfonso},
  title     = {On the concurrent versioning of metamodels and models: challenges and possible solutions},
  year      = {2011},
  isbn      = {9781450306683},
  publisher = {Association for Computing Machinery},
  doi       = {10.1145/2000410.2000414},
  booktitle = {Proceedings of the 2nd International Workshop on Model Comparison in Practice},
  pages     = {16–25},
  numpages  = {10},
  location  = {Zurich, Switzerland},
  series    = {IWMCP '11}
}

@inbook{Cleve2006,
  author    = {Cleve, Anthony
               and Hainaut, Jean-Luc},
  editor    = {L{\"a}mmel, Ralf
               and Saraiva, Jo{\~a}o
               and Visser, Joost},
  title     = {Co-transformations in Database Applications Evolution},
  booktitle = {Generative and Transformational Techniques in Software Engineering: International Summer School, GTTSE 2005, Braga, Portugal, July 4-8, 2005. Revised Papers},
  year      = {2006},
  publisher = {Springer Berlin Heidelberg},
  address   = {Berlin, Heidelberg},
  pages     = {409--421},
  isbn      = {978-3-540-46235-4},
  doi       = {10.1007/11877028_17},
}

@article{curino08,
  author     = {Curino, Carlo A. and Moon, Hyun J. and Zaniolo, Carlo},
  title      = {Graceful database schema evolution: the PRISM workbench},
  year       = {2008},
  issue_date = {August 2008},
  publisher  = {VLDB Endowment},
  volume     = {1},
  number     = {1},
  issn       = {2150-8097},
  doi        = {10.14778/1453856.1453939},
  journal    = {Proceedings VLDB Endowment},
  month      = aug,
  pages      = {761--772},
  numpages   = {12}
}

@online{edgedb,
title = {Schema migrations},
author = {EdgeDB Inc},
url = {https://docs.edgedb.com/guides/migrations},
urldate = {2024-05-01},
year = 2024
}

@inproceedings{Ellis89,
author = {Ellis, C. A. and Gibbs, S. J.},
title = {Concurrency Control in Groupware Systems},
year = {1989},
isbn = {0897913175},
publisher = {Association for Computing Machinery},
doi = {10.1145/67544.66963},
booktitle = {Proceedings of the 1989 ACM SIGMOD International Conference on Management of Data},
pages = {399--407},
numpages = {9},
location = {Portland, Oregon, USA},
series = {SIGMOD '89}
}

@inproceedings{evolvedb,
  author    = {Eckwert, Torben and Guckert, Michael and Taentzer, Gabriele},
  title     = {EvolveDB: a tool for model driven schema evolution},
  year      = {2022},
  isbn      = {9781450394673},
  publisher = {Association for Computing Machinery},
  doi       = {10.1145/3550356.3559095},
  booktitle = {Proceedings of the 25th International Conference on Model Driven Engineering Languages and Systems: Companion Proceedings},
  pages     = {61--65},
  numpages  = {5},
  location  = {Montreal, Quebec, Canada},
  series    = {MODELS '22}
}

@article{Foster2007,
  author     = {Foster, J. Nathan and Greenwald, Michael B. and Moore, Jonathan T. and Pierce, Benjamin C. and Schmitt, Alan},
  title      = {Combinators for bidirectional tree transformations: A linguistic approach to the view-update problem},
  year       = {2007},
  issue_date = {May 2007},
  publisher  = {Association for Computing Machinery},
  address    = {New York, NY, USA},
  volume     = {29},
  number     = {3},
  issn       = {0164-0925},
  doi        = {10.1145/1232420.1232424},
  journal    = {ACM Transsctions on Programming Languages and Systems},
  month      = may,
  pages      = {17–es},
  numpages   = {65}
}

@online{Gemstone,
title={Gemstone Programmer's Guide: Class versions and Instance Migration},
author={GemTalk Systems},
url={https://downloads.gemtalksystems.com/docs/GemStone64/3.2.x/GS64-ProgGuide-3.2/10-ClassHistory.htm},
urldate = {2024-05-01},
year={2015}
}

@inproceedings{perez13,
  author    = {Perez De Rosso, Santiago and Jackson, Daniel},
  title     = {What's wrong with git? a conceptual design analysis},
  year      = {2013},
  isbn      = {9781450324724},
  publisher = {Association for Computing Machinery},
  address   = {New York, NY, USA},
  url       = {https://doi.org/10.1145/2509578.2509584},
  doi       = {10.1145/2509578.2509584},
  booktitle = {Proceedings of the 2013 ACM International Symposium on New Ideas, New Paradigms, and Reflections on Programming \& Software},
  pages     = {37–52},
  numpages  = {16},
  location  = {Indianapolis, Indiana, USA},
  series    = {Onward! 2013}
}

@book{Goldberg80,
author = {Goldberg, Adele},
title = {SMALLTALK-80:  The Interactive Programming Environment},
year = {1984},
isbn = {0201113724},
publisher = {Addison-Wesley Longman Publishing Co., Inc.},
address = {USA}
}

@inproceedings{herrmann15,
  author    = {Herrmann, Kai
               and Voigt, Hannes
               and Behrend, Andreas
               and Lehner, Wolfgang},
  editor    = {Tadeusz, Morzy
               and Valduriez, Patrick
               and Bellatreche, Ladjel},
  title     = {CoDEL -- A Relationally Complete Language for Database Evolution},
  booktitle = {Advances in Databases and Information Systems},
  year      = {2015},
  publisher = {Springer International Publishing},
  address   = {Cham},
  pages     = {63--76},
  isbn      = {978-3-319-23135-8}
}

@inproceedings{herrmann17,
  author    = {Herrmann, Kai and Voigt, Hannes and Behrend, Andreas and Rausch, Jonas and Lehner, Wolfgang},
  title     = {Living in Parallel Realities: Co-Existing Schema Versions with a Bidirectional Database Evolution Language},
  year      = {2017},
  isbn      = {9781450341974},
  publisher = {Association for Computing Machinery},
  address   = {New York, NY, USA},
  doi       = {10.1145/3035918.3064046},
  booktitle = {Proceedings of the 2017 ACM International Conference on Management of Data},
  pages     = {1101–1116},
  numpages  = {16},
  location  = {Chicago, Illinois, USA},
  series    = {SIGMOD '17}
}

@inproceedings{Herrmannsdoerfer11,
  author    = {Herrmannsdoerfer, Markus
               and Vermolen, Sander D.
               and Wachsmuth, Guido},
  editor    = {Malloy, Brian
               and Staab, Steffen
               and van den Brand, Mark},
  title     = {An Extensive Catalog of Operators for the Coupled Evolution of Metamodels and Models},
  booktitle = {Software Language Engineering},
  year      = {2011},
  publisher = {Springer Berlin Heidelberg},
  address   = {Berlin, Heidelberg},
  pages     = {163--182},
  isbn      = {978-3-642-19440-5}
}

@article{JVisser08,
  author     = {Visser, Joost},
  title      = {Coupled Transformation of Schemas, Documents, Queries, and Constraints},
  year       = {2008},
  issue_date = {May, 2008},
  publisher  = {Elsevier Science Publishers B. V.},
  address    = {NLD},
  volume     = {200},
  number     = {3},
  issn       = {1571-0661},
  doi        = {10.1016/j.entcs.2008.04.090},
  journal    = {Electronic Notes Theoretical Computer Science},
  month      = may,
  pages      = {3–23},
  numpages   = {21}
}

@inproceedings{kandel11,
  author    = {Kandel, Sean and Paepcke, Andreas and Hellerstein, Joseph and Heer, Jeffrey},
  title     = {Wrangler: interactive visual specification of data transformation scripts},
  year      = {2011},
  isbn      = {9781450302289},
  publisher = {Association for Computing Machinery},
  address   = {New York, NY, USA},
  doi       = {10.1145/1978942.1979444},
  booktitle = {Proceedings of the SIGCHI Conference on Human Factors in Computing Systems},
  pages     = {3363–3372},
  numpages  = {10},
  location  = {Vancouver, BC, Canada},
  series    = {CHI '11}
}

@inproceedings{diff3,
  address   = {Berlin, Heidelberg},
  author    = {Khanna, Sanjeev and Kunal, Keshav and Pierce, Benjamin C.},
  booktitle = {FSTTCS 2007: Foundations of Software Technology and Theoretical Computer Science},
  editor    = {Arvind, V. and Prasad, Sanjiva},
  isbn      = {978-3-540-77050-3},
  pages     = {485--496},
  publisher = {Springer Berlin Heidelberg},
  title     = {A Formal Investigation of Diff3},
  year      = {2007}
}

@inproceedings{lammel16,
  author    = {L\"{a}mmel, Ralf},
  title     = {Coupled software transformations revisited},
  year      = {2016},
  isbn      = {9781450344470},
  publisher = {Association for Computing Machinery},
  doi       = {10.1145/2997364.2997366},
  booktitle = {Proceedings of the 2016 ACM SIGPLAN International Conference on Software Language Engineering},
  pages     = {239–252},
  numpages  = {14},
  series    = {SLE 2016},
  location  = {Amsterdam, Netherlands}
}

@inproceedings{li99,
  author    = {Xue Li},
  title     = {A Survey of Schema Evolution in Object-Oriented Databases},
  booktitle = {{TOOLS} 1999: 31st International Conference on Technology of Object-Oriented
               Languages and Systems, 22-25 September 1999, Nanjing, China},
  pages     = {362--371},
  publisher = {{IEEE} Computer Society},
  year      = {1999},
  doi       = {10.1109/TOOLS.1999.796507}
}

@inproceedings{Oster06,
  author    = {Oster, Gerald and Molli, Pascal and Urso, Pascal and Imine, Abdessamad},
  booktitle = {2006 International Conference on Collaborative Computing: Networking, Applications and Worksharing},
  title     = {Tombstone Transformation Functions for Ensuring Consistency in Collaborative Editing Systems},
  year      = {2006},
  volume    = {},
  number    = {},
  pages     = {1-10},
  doi       = {10.1109/COLCOM.2006.361867}
}

@article{petersohn20,
  author     = {Petersohn, Devin and Macke, Stephen and Xin, Doris and Ma, William and Lee, Doris and Mo, Xiangxi and Gonzalez, Joseph E. and Hellerstein, Joseph M. and Joseph, Anthony D. and Parameswaran, Aditya},
  title      = {Towards scalable dataframe systems},
  year       = {2020},
  issue_date = {August 2020},
  publisher  = {VLDB Endowment},
  volume     = {13},
  number     = {12},
  issn       = {2150-8097},
  doi        = {10.14778/3407790.3407807},
  journal    = {Procedings VLDB Endowment},
  month      = jul,
  pages      = {2033--2046},
  numpages   = {14}
}

@book{ProGit,
  author    = {Chacon, Scott and Straub, Ben},
  title     = {Pro Git},
  year      = {2014},
  isbn      = {1484200772},
  publisher = {Apress},
  address   = {USA},
  edition   = {2nd},
}

@inproceedings{Ressel96,
  author    = {Ressel, Matthias and Nitsche-Ruhland, Doris and Gunzenh\"{a}user, Rul},
  title     = {An Integrating, Transformation-Oriented Approach to Concurrency Control and Undo in Group Editors},
  year      = {1996},
  isbn      = {0897917650},
  publisher = {Association for Computing Machinery},
  doi       = {10.1145/240080.240305},
  booktitle = {Proceedings of the 1996 ACM Conference on Computer Supported Cooperative Work},
  pages     = {288--297},
  numpages  = {10},
  location  = {Boston, Massachusetts, USA},
  series    = {CSCW '96}
}

@book{sadalage12,
  title     = {NoSQL Distilled: A Brief Guide to the Emerging World of Polyglot Persistence},
  author    = {Sadalage, Pramodkumar J. and Fowler, Martin},
  isbn      = {9780133036121},
  year      = {2012},
  publisher = {Pearson Education}
}

@misc{scherzinger13,
  title         = {Managing Schema Evolution in NoSQL Data Stores},
  author        = {Stefanie Scherzinger and Meike Klettke and Uta St{\"o}rl},
  year          = {2013},
  eprint        = {1308.0514},
  archiveprefix = {arXiv},
  primaryclass  = {cs.DB}
}

@inproceedings{SemanticDeltas,
  author    = {van der Storm, Tijs},
  booktitle = {2013 1st International Workshop on Live Programming (LIVE)},
  title     = {Semantic deltas for live DSL environments},
  year      = {2013},
  volume    = {},
  number    = {},
  pages     = {35-38},
  doi       = {10.1109/LIVE.2013.6617347}
}

@techreport{Shapiro11,
  title       = {{A comprehensive study of Convergent and Commutative Replicated Data Types}},
  author      = {Shapiro, Marc and Pregui{\c c}a, Nuno and Baquero, Carlos and Zawirski, Marek},
  url         = {https://inria.hal.science/inria-00555588},
  type        = {Research Report},
  number      = {RR-7506},
  pages       = {50},
  institution = {{Inria -- Centre Paris-Rocquencourt ; INRIA}},
  year        = {2011},
  month       = Jan,
  hal_id      = {inria-00555588},
  hal_version = {v1}
}

@inproceedings{storl20,
  author = {St{\"o}rl, Uta and Klettke, Meike and Scherzinger, Stefanie},
  year   = {2020},
  month  = {04},
  pages  = {},
  title  = {NoSQL Schema Evolution and Data Migration: State-of-the-Art and Opportunities (Tutorial)},
  booktitle={Proceedings of the 22nd International Conference on Extending Database Technology (EDBT)},
  issn = {2367-2005},
  doi    = {10.5441/002/edbt.2020.87}
}

@inproceedings{storl22,
  author = {St{\"o}rl, Uta and Klettke, Meike},
  year   = {2022},
  month  = {03},
  pages  = {},
  title  = {Darwin: A Data Platform for NoSQL Schema Evolution Management and Data Migration},
  booktitle = {Proceedings of the Workshops of the {EDBT/ICDT} 2022 Joint Conference, Edinburgh, UK, March 29, 2022},
  volume    = {3135},
  publisher = {CEUR-WS.org},
  url = {https://ceur-ws.org/Vol-3135/dataplat_short3.pdf}
}

@article{techdims,
  author    = {Joel Jakubovic and
               Jonathan Edwards and
               Tomas Petricek},
  title     = {Technical Dimensions of Programming Systems},
  journal   = {The Art, Science, and Engineering of Programming},
  volume    = {7},
  number    = {3},
  year      = {2023},
  issn =  {2473-7321},
  doi       = {10.22152/PROGRAMMING-JOURNAL.ORG/2023/7/13},
}

@inproceedings{vermolen11,
  author    = {Vermolen, Sander Dani\"{e}l and Wachsmuth, Guido and Visser, Eelco},
  title     = {Generating database migrations for evolving web applications},
  year      = {2011},
  isbn      = {9781450306898},
  publisher = {Association for Computing Machinery},
  doi       = {10.1145/2047862.2047876},
  booktitle = {Proceedings of the 10th ACM International Conference on Generative Programming and Component Engineering},
  pages     = {83–92},
  numpages  = {10},
  location  = {Portland, Oregon, USA},
  series    = {GPCE '11}
}

@ARTICLE{AIassistants,
  author={Petricek, Tomas and Burg, Gerrit J. J. van den and Nazábal, Alfredo and Ceritli, Taha and Jiménez-Ruiz, Ernesto and Williams, Christopher K. I.},
  journal={IEEE Transactions on Knowledge and Data Engineering},
  title={AI Assistants: A Framework for Semi-Automated Data Wrangling},
  year={2023},
  volume={35},
  number={9},
  pages={9295-9306},
  doi={10.1109/TKDE.2022.3222538}}

@article{challenge-problems,
  title     = {Schema Evolution in Interactive Programming Systems},
  volume    = {9},
  issn      = {2473-7321},
  url       = {http://dx.doi.org/10.22152/programming-journal.org/2025/9/2},
  doi       = {10.22152/programming-journal.org/2025/9/2},
  number    = {1},
  journal   = {The Art, Science, and Engineering of Programming},
  publisher = {Aspect-Oriented Software Association (AOSA)},
  author    = {Edwards, Jonathan and Petricek, Tomas and van der Storm, Tijs and Litt, Geoffrey},
  year      = {2024},
  month     = oct
}

@inproceedings{denicek,
  author    = {Petricek, Tomas and Edwards, Jonathan},
  title     = {Denicek: Computational Substrate for Document-Oriented End-User Programming},
  year      = {2025},
  isbn      = {9798400720376},
  publisher = {Association for Computing Machinery},
  address   = {New York, NY, USA},
  url       = {https://doi.org/10.1145/3746059.3747646},
  doi       = {10.1145/3746059.3747646},
  booktitle = {Proceedings of the 38th Annual ACM Symposium on User Interface Software and Technology},
  articleno = {32},
  numpages  = {19},
  location  = {},
  series    = {UIST '25}
}

@inproceedings{sculpin,
  author    = {Horowitz, Joshua and Hayatpur, Devamardeep and Xia, Haijun and Heer, Jeffrey},
  title     = {Sculpin: Direct-Manipulation Transformation of JSON},
  year      = {2025},
  isbn      = {9798400720376},
  publisher = {Association for Computing Machinery},
  address   = {New York, NY, USA},
  url       = {https://doi.org/10.1145/3746059.3747651},
  doi       = {10.1145/3746059.3747651},
  booktitle = {Proceedings of the 38th Annual ACM Symposium on User Interface Software and Technology},
  articleno = {34},
  numpages  = {15},
  location  = {},
  series    = {UIST '25}
}

@inproceedings{church2014case,
  title     = {A case of computational thinking: The subtle effect of hidden dependencies on the user experience of version control.},
  author    = {Church, Luke and S{\"o}derberg, Emma and Elango, Elayabharath},
  booktitle = {PPIG},
  pages     = {16},
  year      = {2014}
}

@misc{xkcd1597,
  title   = {Git},
  url     = {https://xkcd.com/1597/},
  journal = {xkcd},
  author  = {Munroe, Randall},
    note   = {[Online; accessed 29-August-2025]}
}

@misc{philomatics-git,
  title   = {Never use git cherry-pick},
  url     = {https://www.youtube.com/watch?v=WPCxtFkLa7g},
  author  = {von Franqué, Alexander},
    note   = {[Online; accessed 29-August-2025]}
}

@inproceedings{edit-lenses,
  author    = {Hofmann, Martin and Pierce, Benjamin and Wagner, Daniel},
  title     = {Edit lenses},
  year      = {2012},
  isbn      = {9781450310833},
  publisher = {Association for Computing Machinery},
  address   = {New York, NY, USA},
  url       = {https://doi.org/10.1145/2103656.2103715},
  doi       = {10.1145/2103656.2103715},
  booktitle = {Proceedings of the 39th Annual ACM SIGPLAN-SIGACT Symposium on Principles of Programming Languages},
  pages     = {495–508},
  numpages  = {14},
  location  = {Philadelphia, PA, USA},
  series    = {POPL '12}
}

@book{cypher93-pbd,
  title={Watch what I do: {P}rogramming by demonstration},
  author={Cypher, Allen and Halbert, Daniel Conrad},
  year={1993},
  publisher={MIT press}
}

@misc{miller-columns,
  author       = {{Wikipedia contributors}},
  title        = {Miller columns --- {Wikipedia}{,} The Free Encyclopedia},
  year         = {2025},
  howpublished = {\url{https://en.wikipedia.org/w/index.php?title=Miller_columns&oldid=1305075621}},
  note         = {[Online; accessed 12-September-2025]}
}

@phdthesis{bakke:phdthesis,
  sortname = {Bakke2016-09},
  author   = {Eirik Bakke},
  title    = {Expressive Query Construction through Direct Manipulation of Nested Relational Results},
  school   = {Massachusetts Institute of Technology},
  year     = {2016},
  month    = {9},
  note     = {Available at \url{https://dspace.mit.edu/handle/1721.1/107280}}
}

@online{ultorg,
  author  = {Eirik Bakke},
  title   = {Ultorg},
  year    = 2025,
  url     = {https://www.ultorg.com},
  urldate = {2025-09-14}
}

@inproceedings{sun00,
  author    = {Sun, Chengzheng},
  title     = {Undo any operation at any time in group editors},
  year      = {2000},
  isbn      = {1581132220},
  publisher = {Association for Computing Machinery},
  address   = {New York, NY, USA},
  url       = {https://doi.org/10.1145/358916.358990},
  doi       = {10.1145/358916.358990},
  booktitle = {Proceedings of the 2000 ACM Conference on Computer Supported Cooperative Work},
  pages     = {191–200},
  numpages  = {10},
  location  = {Philadelphia, Pennsylvania, USA},
  series    = {CSCW '00}
}

@techreport{weiss08,
  title       = {{A Flexible Undo Framework for Collaborative Editing}},
  author      = {Weiss, St{\'e}phane and Urso, Pascal and Molli, Pascal},
  url         = {https://inria.hal.science/inria-00275754},
  type        = {Research Report},
  number      = {RR-6516},
  institution = {{INRIA}},
  year        = {2008},
  hal_id      = {inria-00275754},
  hal_version = {v2}
}

@article{berlage94,
  author     = {Berlage, Thomas},
  title      = {A selective undo mechanism for graphical user interfaces based on command objects},
  year       = {1994},
  issue_date = {Sept. 1994},
  publisher  = {Association for Computing Machinery},
  address    = {New York, NY, USA},
  volume     = {1},
  number     = {3},
  issn       = {1073-0516},
  url        = {https://doi.org/10.1145/196699.196721},
  doi        = {10.1145/196699.196721},
  journal    = {ACM Trans. Comput.-Hum. Interact.},
  month      = sep,
  pages      = {269–294},
  numpages   = {26}
}

@online{braid,
  author  = {Michael Toomim},
  title   = {Collapsing Time Machines},
  year    = 2025,
  url     = {https://braid.org/time-machines},
  urldate = {2025-09-14}
}

@online{patchwork,
  author  = {Geoffrey Litt and
Paul Sonnentag and
Max Schöning and
Adam Wiggins and
Peter van Hardenberg and
Orion Henry},
  title   = {Patchwork},
  year    = 2024,
  url     = {https://www.inkandswitch.com/patchwork/notebook/},
  urldate = {2025-09-14}
}

@online{ceda,
  author  = {David Barrett-Lennard},
  title   = {Introducing CEDA},
  year    = 2010,
  url     = {https://cedanet.com.au/ceda/papers/Introducing%20CEDA.pdf},
  urldate = {2025-09-14}
}

@online{ceda-ot,
  author  = {David Barrett-Lennard},
  title   = {Operational Transformation},
  year    = 2010,
  url     = {https://cedanet.com.au/ceda/ot/},
  urldate = {2025-09-14}
}

@inproceedings{czarnecki2009bidirectional,
  author    = {Czarnecki, Krzysztof
               and Foster, J. Nathan
               and Hu, Zhenjiang
               and L{\"a}mmel, Ralf
               and Sch{\"u}rr, Andy
               and Terwilliger, James F.},
  editor    = {Paige, Richard F.},
  title     = {Bidirectional Transformations: A Cross-Discipline Perspective},
  booktitle = {Theory and Practice of Model Transformations},
  year      = {2009},
  publisher = {Springer Berlin Heidelberg},
  address   = {Berlin, Heidelberg},
  pages     = {260--283},
  isbn      = {978-3-642-02408-5}
}

@book{hu2024bidirectional,
  editor    = {Hu, Zhenjiang and Onizuka, Makoto and Yoshikawa, Masatoshi},
  title     = {Bidirectional Collaborative Data Management: Collaboration Frameworks for Decentralized Systems},
  publisher = {Springer},
  year      = {2024},
  isbn      = {978-981-97-6428-0}
}

@inproceedings{ignat-synql,
  title       = {{Synql: A CRDT-Based Approach for~Replicated Relational Databases with~Integrity Constraints}},
  author      = {Ignat, Claudia-Lavinia and Elvinger, Victorien and Ba, Habibatou},
  url         = {https://inria.hal.science/hal-04969158},
  booktitle   = {{Lecture Notes in Computer Science}},
  address     = {Groningen, Netherlands},
  editor      = {Rolando Martins},
  publisher   = {{Springer Nature Switzerland}},
  series      = {Distributed Applications and Interoperable Systems},
  volume      = {LNCS-14677},
  pages       = {18-35},
  year        = {2024},
  month       = Jun,
  doi         = {10.1007/978-3-031-62638-8\_2},
  hal_id      = {hal-04969158},
  hal_version = {v4}
}

@article{exelmans25,
  author        = {Exelmans, Joeri and Teodorov, Ciprian and Vangheluwe, Hans},
  date          = {2025/06/01},
  doi           = {10.1007/s10270-024-01212-x},
  id            = {Exelmans2025},
  isbn          = {1619-1374},
  journal       = {Software and Systems Modeling},
  number        = {3},
  pages         = {721--739},
  title         = {Operation-based versioning as a foundation for live executable models},
  url           = {https://doi.org/10.1007/s10270-024-01212-x},
  volume        = {24},
  year          = {2025}
}

@inproceedings{carvalho24,
  author    = {Carvalho, Lu{\'\i}s and Costa Seco, Jo\~{a}o},
  title     = {{A Language-Based Version Control System for Python}},
  booktitle = {38th European Conference on Object-Oriented Programming (ECOOP 2024)},
  pages     = {9:1--9:27},
  series    = {Leibniz International Proceedings in Informatics (LIPIcs)},
  isbn      = {978-3-95977-341-6},
  issn      = {1868-8969},
  year      = {2024},
  volume    = {313},
  editor    = {Aldrich, Jonathan and Salvaneschi, Guido},
  publisher = {Schloss Dagstuhl -- Leibniz-Zentrum f{\"u}r Informatik},
  address   = {Dagstuhl, Germany},
  url       = {https://drops.dagstuhl.de/entities/document/10.4230/LIPIcs.ECOOP.2024.9},
  urn       = {urn:nbn:de:0030-drops-208586},
  doi       = {10.4230/LIPIcs.ECOOP.2024.9}
}

@online{automerge-conflicts,
  author  = {Automerge Contributors},
  title   = {Automerge Conflicts},
  year    = 2025,
  url     = {https://automerge.org/docs/reference/documents/conflicts/},
  urldate = {2025-09-14}
}

@online{couchdb-conflicts,
  author  = {Apache CouchDB Contributors},
  title   = {CouchDB Replication and Conflict Model},
  year    = 2025,
  url     = {https://docs.couchdb.org/en/stable/replication/conflicts.html#},
  urldate = {2025-09-14}
}

@article{kamina25,
  author    = {Tetsuo Kamina and Tomoyuki Aotani and Hidehiko Masuhara},
  title     = {Evolution Language Framework for Persistent Objects},
  journal   = {The Art, Science, and Engineering of Programming},
  year      = 2025,
  volume    = 10,
  number    = 1,
  month     = feb,
  doi       = {10.22152/programming-journal.org/2025/10/12},
  date      = {2025-02-15},
  submitted = {2024-10-01},
  url       = {https://2025.programming-conference.org/}
}

\end{document}